\documentclass[prd,twocolumn,showpacs,nofootinbib]{revtex4}
\usepackage{subfigure}
\usepackage{amssymb}
\usepackage{epsfig}
\usepackage{float}
\usepackage{caption}
\usepackage{color}

\newcommand{\be}{\begin{equation}}
\newcommand{\ee}{\end{equation}}
\newcommand{\bea}{\begin{eqnarray}}
\newcommand{\eea}{\end{eqnarray}}
\newcommand{\C}{\mathcal{C}}
\newcommand{\I}{\mathcal{I}}

\input{tcilatex}

\begin{document}

\title{\hspace{75pt} Cosmology on all scales: \newline a two-parameter perturbation expansion}

\author{Sophia R. Goldberg}
\email{s.r.goldberg@qmul.ac.uk}
\affiliation{School of Physics and Astronomy, Queen Mary University of London, Mile End Road, London E1 4NS, UK.}

\author{Timothy Clifton}
\email{t.clifton@qmul.ac.uk}
\affiliation{School of Physics and Astronomy, Queen Mary University of London, Mile End Road, London E1 4NS, UK.}

\author{Karim A. Malik}
\email{k.malik@qmul.ac.uk}
\affiliation{School of Physics and Astronomy, Queen Mary University of London, Mile End Road, London E1 4NS, UK.}

\begin{abstract}
We propose and construct a two-parameter perturbative expansion around a Friedmann-Lema\^{i}tre-Robertson-Walker geometry that can be used to model high-order gravitational effects in the presence of non-linear structure. This framework reduces to the weak-field and slow-motion post-Newtonian treatment of gravity in the appropriate limits, but also includes the low-amplitude large-scale fluctuations that are important for cosmological modelling. We derive a set of field equations that can be applied to the late Universe, where non-linear structure exists on supercluster scales, and perform a detailed investigation of the associated gauge problem. This allows us to identify a consistent set of perturbed quantities in both the gravitational and matter sectors, and to construct a set of gauge-invariant quantities that correspond to each of them. The field equations, written in terms of these quantities, take on a relatively simple form, and allow the effects of small-scale structure on the large-scale properties of the Universe to be clearly identified. We find that inhomogeneous structures source the global expansion, that there exist new field equations at new orders, and that there is vector gravitational potential that is a hundred times larger than one might naively expect from cosmological perturbation theory. Finally, we expect our formalism to be of use for calculating relativistic effects in upcoming ultra-large-scale surveys, as the form of the gravitational coupling between small and large scales depends on the non-linearity of Einstein's equations, and occurs at what is normally thought of as first order in cosmological perturbations.
\end{abstract}

\maketitle


\section{Introduction} \label{intro}

A crucial feature of our observable Universe is that it contains many gravitationally-bound structures, on a variety of different scales. These range from stars and planets to the galaxies, clusters and superclusters that make up the cosmic web we observe today. A challenge for theoretical cosmologists is how to consistently model this array of structures, given that their density contrasts can be very large, and that we wish to consider distance scales as large as the Hubble radius. The standard approach to this problem is to assume a global Friedmann-Lema\^{i}tre-Robertson-Walker (FLRW) background, and to use a mixture of cosmological perturbation theory and Newtonian gravity in order to model the effects of additional weak gravitational fields, see {\it e.g.} \cite{malik, RDtextbook, MillSim}. This approach works extremely well for a wide variety of situations, but it starts to become problematic when one tries to consider non-linear relativistic gravity. This is because non-linear density contrasts do not naturally fit into the formalism of cosmological perturbation theory, and because on small scales the velocity of matter and the gradients of gravitational potentials can both be large.

Our approach to addressing this problem is to simultaneously expand the metric and energy-momentum tensor using both cosmological and post-Newtonian perturbation theories \cite{will, poisson}. The result of this can formally be described as a perturbative expansion in two parameters, which we expect to be a consistent and valid description of both non-linear structure on small scales and linear fluctuations on horizon-sized scales. Such a formalism therefore enables one to model the effects of non-linear structure on the dynamics of large-scale cosmological perturbations, as well as on the cosmological background itself. It provides a more representative picture of the real Universe than either cosmological perturbation theory or post-Newtonian theory could by themselves, and may be of use for consistently modelling the relativistic effects that future surveys will seek to detect.

The reason that standard cosmological perturbation theory is not ideal for modelling structure on scales less than about $100$Mpc (in the late Universe) is that below this scale both the density contrasts and velocities start to become large, in comparison to the background energy density and gravitational potentials. Moreover, perturbations to the metric appear at the same order in the field equations as terms that are as large as the dynamical background. This has led to much study of the idea that the formation of structure in the Universe could have a strong ``back-reaction'' effect on the large-scale expansion, as the perturbative expansion itself may start to break down \cite{chrisBack, BRCPTbreaks, vdhogen, BackClifton, Buchert1, Buchert2, Ellis}. Although many authors now believe the effects of back-reaction on the FLRW background to be small, this does not necessarily mean that the effects of small-scale structure on large-scale cosmological perturbations must also be small. To address this latter question requires an approach that can systematically and consistently track the effect of non-linear structures order-by-order in perturbation theory, which is exactly what our two-parameter perturbative expansion is designed to do.

In some respects, our treatment of the gravity on small scales resembles the quasi-static (or slow-motion) limit of cosmological perturbation theory. This approach has often been used in the literature to describe small-scale structure \cite{Peebles}, and, at lowest order, gives a set of equations that look a lot like those of Newtonian gravity. The basic idea in this approach is to neglect terms in the field equations that involve time derivatives, as these are generally expected to be small in comparison to spatial derivatives. What is unclear in the usual application of the quasi-static limit is how this approach can be extended to non-linear gravity. The terms involving time derivatives that were discarded may or may not appear at next order in perturbations, and it may or may not be necessary to adjust the order-of-smallness of velocities or vector potentials in order to make the entire system of equations consistent. The post-Newtonian expansion that we employ could, in some sense, be viewed as a formalised version of the quasi-static limit of cosmological perturbation theory, as it consistently tracks the smallness of time derivatives, and the consequences that result from the smallness of time derivatives.

Of course, one of the main application of constructing a perturbative expansion of the type outlined above is to determine the signatures of Einstein's theory in cosmological data. Studies with this goal have already been performed using second-order cosmological perturbation theory \cite{Bonvin, Bonvin2, Bonvin3, Bruni:2014xma, BenDayan:2012ct, BenDayan:2013gc}, and we expect it to be a matter of significant interest to determine whether a framework that formalizes the quasi-static limit can be used to simplify or extend them. Hints that this should be possible come from studies of second-order gravitational fields that average to the size of first order fields \cite{Alan, Chris2, Kolb1, Rasanen, Adamek}, and calculations that suggest the second-order vector potential to be a hundred times larger than naively expected \cite{AndEtAl}. Both of these turn out to be natural results of the formalism we present in this paper, which may therefore prove useful for gaining a full understanding of the results from upcoming high-precision surveys \cite{euclid, SKA, LSST}. 

In the case where long-wavelength cosmological perturbations are neglected, our formalism reduces to post-Newtonian gravity on an expanding background \cite{bruni, RDformalism, RDnature, AdamekFormalism}. If the scale of the post-Newtonian system is small enough, then the background expansion only influences the local physics of that system at high orders in perturbation theory. This means we end up with a set of equations that are consistent with post-Newtonian gravity up to the accuracy of current observations but which differ to post-Newtonian gravity at high-order. Our framework could therefore be used to quantify the effects of cosmological expansion and cosmological potentials on weak-field systems, if this was ever required. It could also be used to formally model gravitational fields in relativistic N-body simulations \cite{RDgevolution, bruniNbody, Nbodycor, NvR, GreenWald, Chisari, Fidler, RDnature, Adamek:2015hqa, Adamek:2014xba, Christopherson:2015ank}, and the effects of small scale fluctuations on cosmological observables such as galaxy number counts \cite{galNoRD, Bonvin}. Additionally, such a theory has the potential to offer new ways of testing Einstein's theory. For most of what follows, in this paper, we will take the weak-field systems being modelled to be clusters and superclusters.

In Section \ref{sec:pertexp} we introduce the relevant perturbative expansions for our formalism: post-Newtonian gravity, cosmological perturbation theory and our two-parameter expansion. In Section \ref{obsjust} we consider, using observational results, the size of quantities such as gravitational potentials and energy densities for various physical systems. This indicates which of the two perturbative expansions we should expect to apply to each system. In Section \ref{sec:fe} we use our two-parameter expansion to derive the field equations that correspond to structure on supercluster scales. The expressions that result are lengthy, so in Section \ref{Gchoice} we define a two-parameter coordinate transformation that can be applied to the metric and stress-energy tensor. This enables us to construct gauge-invariant quantities in Section \ref{sec:gaugeinvariants}, and to write gauge-invariant versions of the field equations. This simplifies the field equations, and allows us to determine at which orders we should expect perturbations to appear. In Section \ref{SmallAndLarge} we discuss our final field equations, and consider how our formalism could be applied to the smallest and largest gravitationally-bound structures that exist in the Universe. Finally, we conclude in Section \ref{conc}.

We use Latin and Greek indices to denote space and space-time indices, respectively. Commas and dots denote the partial derivatives and derivatives with respect to coordinate time $t$, respectively, such that
\bea
f_{, \mu} \equiv \frac{\partial f}{\partial x^{\mu}} \, , \qquad \dot{f} \equiv  \frac{\partial f}{\partial t} \, ,  \nonumber 
\eea
where $x^{\mu}$ are space-time coordinates and $f$ is any function on space-time. Additionally, we choose units such that $c =G = 1$, so that Einstein's field equations are given by 
\be
\label{EE}
R_{\mu \nu} = 8 \pi \left(T_{\mu \nu} - {\textstyle\frac{1}{2}}T g_{\mu \nu}\right)\, ,
\ee
where $R_{\mu \nu}$ is the Ricci tensor of the space-time metric $g_{\mu \nu}$, where $T_{\mu \nu}$ is the energy-momentum tensor of the matter fields within the space-time, and where $T \equiv T^{\mu}_{\mu}$. Throughout this paper we will treat the matter fields as a perfect fluid, so that the energy-momentum tensor can be written as
\be \label{e:Tmunu}
T_{\mu \nu} = (\rho + p) u_{\mu} u_{\nu} + p g_{\mu \nu} \, , 
\ee
where $\rho$ and $p$ are the energy density and pressure measured by observers following four-velocity $u^{\mu } \equiv {dx^{\mu}}/{d \tau}$, and where $\tau$ is the proper time comoving with the fluid.

\section{Perturbative expansions }
\label{sec:pertexp}

Perturbative expansions are used extensively in gravitational physics, as the full Einstein equations, given in Eq. (\ref{EE}), are otherwise very difficult to solve. These expansions come in a variety of different forms, and are usually constructed or adapted to be used in specific scenarios that are of particular physical interest. The two perturbative expansions that we wish to use in this paper are the post-Newtonian expansion, and the cosmological perturbation theory expansion. These are by no means the only perturbative constructions that can be applied to understand relativistic gravity, but they are probably the best suited to understanding it in cosmology.

We will start this section by discussing both post-Newtonian and cosmological perturbation theory expansions separately, before moving on to show how each of them needs to be altered from their canonical forms if they are to be used simultaneously to describe astrophysical structures that span the full range, from galaxies all the way through to super-horizon fluctuations. By considering these two expansions simultaneously we will shed light on the link between the gravitational fields of highly non-linear virialized objects, and the large-scale properties of the Universe. These links, and the interplay between gravitational physics on small and large scales, will become increasingly important as we move to higher orders in perturbation theory.

The starting point for both of these expansions is the realisation that the Einstein equations can be written as a set of wave equations, which take the form \cite{wave} 
\be \label{wave}
\square \psi = - 4 \pi \mu \, ,
\ee
where $\square$ is the D'Alembertian operator associated with the metric of space-time, $\psi$ represents the various gravitational potentials associated with the metric, and $\mu$ is a source term (derived from the components of the energy-momentum tensor, and the components of the metric with up to one derivative).

Eq. (\ref{wave}) is a wave equation with null characteristics, so its retarded solutions, assuming certain boundary conditions, are given by integrals of the form
\be
\label{wavesol}
\psi (t, {\bf x}) = \int_\mathcal{C_-} \frac{\mu(t- \vert {\bf x}- {\bf x^{\prime}}\vert, {\bf x^{\prime}})}{\vert {\bf x}- {\bf x^{\prime}}\vert} d^3 x^{\prime} \, ,
\ee
where $\mathcal{C_-}$ the past light cone of the point $x=(t, {\bf x})$. These solutions, in general, represent a set of waves, with a characteristic wavelength and frequency that are determined by the source, $\mu$. We will refer to these as $\lambda_c$ and $\omega_c$, respectively. Because Eq. (\ref{wavesol}) represents a set of null waves, these quantities are related by $\lambda_c=2 \pi /\omega_c$.

So far, we have not used perturbation theory at all. If we want to do this, in order to get concrete solutions to Einstein's equations, then we need to understand how the integral in Eq. (\ref{wavesol}) behaves under the relevant approximations. Specifically, we need to know if the length scale under investigation is smaller or greater than $\lambda_c$. These regimes are often referred to in the relativistic astrophysics literature as the ``near zone'' and the ``wave zone'', respectively \cite{poisson}. We will use the same ideas, but apply them to cosmology instead. We will then refer to these two regimes as the ``Newtonian'' and the ``cosmological''. The relevant expansion for the Newtonian regime will be an adapted version of the post-Newtonian expansion, while the one relevant for the cosmological regime will be an adapted version of cosmological perturbation theory. Let us now consider each of these regimes in turn, before considering them both together.

\subsection{Post-Newtonian gravity}
\label{sec:pn}

In the Newtonian regime (our version of the near zone) we will assume that distance scales are small compared to the characteristic wavelength, $\lambda_c$, such that
\be
L_N \ll \lambda_c = \frac{2 \pi }{\omega_c} =  t_c \, ,
\ee
where we have introduced the characteristic time-scale $t_c$, and the typical length scale associated with the Newtonian regime, $L_N$. Another way of stating this condition would be to say that the velocities of the sources are, in some sense, slow. This follows from the fact that characteristic dimensionless velocities are of the order $v \sim L_N/t_c \ll 1$ (recall that we are using units in which $c=1$). In this sense, small scales tend to correspond to slow motions.

Now consider the consequences of the assumption of small scales for derivatives of the source term, $\mu$. Spatial derivatives are of the order $\vert \nabla \mu \vert \sim \mu / L_N$, while time derivatives are of order $\dot{\mu} \sim \mu/t_c$. We therefore have
\be\label{smallmu}
\dot{\mu} \ll  \vert \nabla \mu \vert \, .
\ee 
In words, the typical variation of the sources in time is small compared to their variation in space. It is also apparent that the order of this smallness should be expected to be of the same size as the dimensionless velocity, $v$. 

Let us now consider the size of the gravitational potentials that are represented by $\psi$, and how they vary in space and time. It is apparent from Eq. (\ref{wavesol}) that if $L_N \sim \vert {\bf x}- {\bf x^{\prime}}\vert \ll t$, and if we Taylor expand the time-dependent part of the integrand, then the leading-order part of $\psi$ is given by
\be\label{solnpsi}
\psi = \int_{\mathcal{V}} \frac{\mu(t,{\bf x^{\prime}})}{\vert {\bf x}- {\bf x^{\prime}}\vert} d^3 x^{\prime} \, ,
\ee 
where $\mathcal{V}$ denotes a space-like volume of constant time. It can now be seen from Eqs. (\ref{smallmu}) and (\ref{solnpsi}) that when $\vert {\bf x}- {\bf x^{\prime}}\vert  \ll t_c$ the derivatives of $\psi$ satisfy \cite{poisson}
\be \label{smallderivs}
\dot{\psi} \ll  \vert \nabla \psi \vert \, .
\ee
Again, the order of smallness of the time derivative, compared to the space derivatives, is found to be of the order of $v$. It can also be seen that $\psi \sim \mu L_N^2$. 

The discussion above all follows from the assumption of small scales (and hence low velocities), as well as the null characteristics of the Einstein field equations. A further requirement to define the post-Newtonian expansion is that the magnitude of the gravitational potentials are themselves small. This point is complicated by the fact that there are a number of gravitational potentials in Einstein's theory, and not just the one that was used for schematic purposes in Eq. (\ref{wave}). The magnitude of a potential depends, through the field equations, on the sources that generate it. The magnitude of any given potential can also be linked to the velocity of the matter fields in the space-time through the equations of motion of those fields. Let us now consider how this works for the leading-order parts of each of the components of the metric. In order to do this, it is convenient to define the parameter
\begin{equation} \label{eta}
\eta \sim v  \sim \frac{\vert \partial/\partial t \vert}{\vert \partial/\partial x \vert}\, ,
\end{equation}
which can be used to keep track of the order-of-smallness of a quantity within this expansion.

At leading order, the space components of the equation of motion for freely falling time-like particles tells us that $\dot{v} \sim \vert \nabla g_{00} \vert$, which implies the metric is perturbed in the following way
\be
g_{00} = g_{00}^{(0)}(t) + g_{00}^{(2)} (t,{\bf x}) + \ldots \, .
\ee
Here we have now used a superscript in brackets to denote the order of a quantity in $\eta$, and where the ellipsis denote terms that are smaller than $\eta^2$. There can be no terms that depend on spatial position at order $\eta$ or larger, as this would be incompatible with the leading-order part of the equation of motion.

Meanwhile, the leading-order part of the time-time component of the field equations gives
\be
\label{newton}
\nabla^2 g_{00} \sim \rho \, ,
\ee
where $\rho$ is the leading-order part of the energy density of the matter fields. This tells us that the $\rho$, which actually corresponds to the mass density, can be no larger than $\eta^2 L_N^{-2}$. The similarity between Eq. (\ref{newton}) and the Newton-Poisson equation also justifies associating $g_{00}^{(2)} (t,{\bf x})$ with the Newtonian gravitational potential, $U$. Furthermore, for freely falling time-like particles we find
\bea
U \sim v^2 \, . \label{Usim2}
\eea

To go to higher-order in $g_{00}$, and to find the leading-order parts of the other components of the metric, we need to consider the higher-order parts of the energy-momentum tensor. To do this we first expand the energy density and pressure as $\rho = \rho^{(2)} + \rho^{(4)} + \ldots$ and $p=p^{(4)} + \ldots$, respectively. The components of the tensor given in Eq. (\ref{e:Tmunu}), up to $\mathcal{O}(\eta^5L_N^{-2})$, are then
\bea\label{t4a}
T_{00}^{(2)} &=&- g_{00}^{(0)} \rho^{(2)}  \\[5pt]
T_{00}^{(4)} &=&- g_{00}^{(0)} \rho^{(4)} -  \rho^{(2)} \left(g_{00}^{(0)}   u^{(1)i} u^{(1)}_i  + g_{00}^{(2)} \right) \\[5pt]
T_{0i}^{(3)} &=& -\sqrt{-g^{(0)}_{00}}\rho^{(2)} u^{(1)}_i \\[5pt]
T_{ij}^{(4)} &=& \rho^{(2)} u^{(1)}_i u^{(1)}_j + p^{(4)} g^{(0)}_{ij}  , \label{t4c}
\eea
where the spatial part of the four-velocity is such that $v \equiv \vert v^{(1)i} \vert \sim \vert u^{(1)i} \vert  \sim \eta$ and we assume $g^{(0)}_{0i} = 0$. In each of these expressions we have continued the practice of using superscripts in brackets to denote the order-of-smallness of a quantity. However, when a quantity is dimensionful, such as $p^{(4)}$, then the reader should take this to mean, for example, $p^{(4)} \sim \eta^4 L_N^{-2}$. 

The post-Newtonian gravitational fields that result from Eqs. (\ref{t4a})-(\ref{t4c}) are then given by
\bea
g_{00} &=& g_{00}^{(0)}(t) + g_{00}^{(2)} (t,{\bf x}) + {\textstyle\frac{1}{2}} g_{00}^{(4)} (t,{\bf x}) \ldots \quad\\[5pt]
g_{ij} &=& g_{ij}^{(0)}(t) + g_{ij}^{(2)} (t,{\bf x}) + \ldots \\[5pt]
g_{0i} &=& g_{0i}^{(3)} (t,{\bf x}) + \ldots\, ,
\eea
where we have assumed that coordinates can be chosen such that $g_{0i}^{(0)}$ vanishes. The metric components $g_{00}^{(4)}$, $g_{ij}^{(2)}$, and $g_{0i}^{(3)}$ are usually referred to as ``post-Newtonian potentials''.

One may note that the first spatially dependent term in $g_{0i}$ occurs at $O(v^3)$. This is because the first non-zero source term for this potential is order $\rho^{(2)} v^{(1)}_i$. It can also be noted that the orders of the gravitational potentials required for them to be labelled ``post-Newtonian'' are different in different parts of the metric. This is because time derivatives add an order-of-smallness, compared to space derivatives, and because these two types of derivatives operate on different components of the metric in the equations of motion of time-like particles.

One may also note that there are a number of missing terms in both the energy-momentum tensor and the metric. For example, there are no terms in $T_{00}$ of $\mathcal{O}\left(\eta^3 L_N^{-2}\right)$, and no terms in $g_{00}$ of $\mathcal{O}(\eta^3)$. As far as the energy-momentum tensor is concerned, this can be considered a choice of the type of matter that one wishes to model. For example, matter with a pressure term at $\mathcal{O}(\eta^2 L_N^{-2})$ could be included, if required, as was recently done in \cite{sanghai2}. One could also include heat flow or anisotropic pressure, if they were required. The situation with the metric, however, is quite different. 

The required order-of-smallness of the different components of the metric is not specified from the outset. It is determined by solving the field equations, and by using the equations of motion of the matter fields. This means that one could, for example, have tried to include a $g_{00}^{(3)}$ term in the time-time component of the metric. However, there would be no matter fields to source such a term, and so it would end up satisfying a homogeneous version of the equation satisfied by $g_{00}^{(2)}$. This means that the hypothesized $g_{00}^{(3)}$ term describes no new physics, and can be absorbed into $g_{00}^{(2)}$ without loss of generality, and it is not necessary or helpful to consider such a term independently. We will return to this point, later on.

After all of this, we therefore end up with a metric and an energy-momentum tensor that are expanded at even orders in $\eta$ in their time-time and space-space components, and at odd orders in $\eta$ in their time-space components (a trend that continues until gravitational waves are generated). We also have that time derivatives add an extra order of smallness to any quantity that they act upon, when compared with space derivatives, and that the lowest-order gravitational potentials are at either $\mathcal{O}(\eta^2)$ or $\mathcal{O}(\eta^3)$. This is all very different to the results of the expansion used in cosmological perturbation theory, which we will review in the next section. For further details about post-Newtonian expansions the reader is referred to the textbooks by Will \cite{will} and Poisson \& Will \cite{poisson}.

\subsection{Cosmological perturbation theory} \label{pertCPT}

Cosmological perturbation theory applies to large scales, up to and beyond the particle horizon of the observable Universe. Such length scales are, by definition, comparable to the characteristic wavelength, $\lambda_c$, such that
\be
L_C \sim \lambda_c = \frac{2 \pi }{\omega_c} =  t_c \, ,
\ee
where $L_C$ is the typical length scale associated with the regime of cosmological perturbation theory. This means that characteristic velocities, $v \sim L_C/t_c$, are not small, and that the variation in time of gravitational potentials and matter fields cannot be considered small when compared to their variation in space.

These facts mean that, unlike the case of post-Newtonian gravity, we cannot use $v$ to track the smallness of gravitational potentials or matter fields. Instead we have to hypothesize, or construct \cite{cwoa, sanghai1}, a global solution to Einstein's equations that can be used as a background to perturb around. For most purposes this is taken to be the FLRW geometry:
\be
\label{flrw}
ds^2 = -dt^2 + a^2(t) \left( \frac{dr^2}{1-k r^2} +r^2 (d\theta^2+ \sin^2 \theta d\phi^2) \right) \, ,
\ee
where $a(t)$ is the scale factor, and $k$ is the curvature of a spatial volume of constant $t$. The precise functional form of $a(t)$ depends on the matter content of the space-time, and the value of the curvature constant, $k$. For the majority of this paper we will consider background geometries in which $k=0$.

With the flat FLRW background in hand, one can now consider small fluctuations to both the metric of space-time and the matter fields that exist within it. Starting with the metric, we can write
\be
g_{\mu \nu} = g_{\mu \nu}^{(0)} (t) + g_{\mu \nu}^{(1)}(t,{\bf x}) + \ldots \, ,
\ee
where $g_{\mu \nu}^{(0)} (t)$ corresponds to the FLRW background, see Eq. (\ref{flrw}), and $g_{\mu \nu}^{(1)}(t,{\bf x})$ corresponds to the leading-order perturbation. These contributions to the metric have, to date, been the only ones required to calculate the vast majority of cosmological gravitational phenomena. The ellipsis in this equation denote terms that are smaller than $g_{\mu \nu}^{(1)}$, and the superscripts in brackets are now being used to denote the order of smallness of a quantity in cosmological perturbation theory (they should not be confused with quantities perturbed in the post-Newtonian expansion, as outlined in the previous section).

If we now perturb the matter fields, then we can write the energy density and pressure within the space-time as $\rho= \rho^{(0)}+\rho^{(1)}+\ldots$ and $p=p^{(0)}+p^{(1)}+\ldots$, where the quantities $\rho^{(0)}$ and $p^{(0)}$ should be understood to be the values of the energy density and pressure in the background FLRW geometry, respectively. Using this, together with the perturbed metric, the components of the energy-momentum tensor can be written to linear order:
\bea
T_{00} &=&  \rho^{(0)} + \rho^{(1)} - g_{00}^{(1)} \rho^{(0)} + \ldots\\[5pt]
T_{0i} &=& - \rho^{(0)}( v_i^{(1)} +g^{(1)}_{0i}) -p^{(0)} v_{i}^{(1)} +\ldots \\[5pt]
T_{ij} &=&  p^{(0)}(g^{(0)}_{ij} +g^{(1)}_{ij}) +p^{(1)}g_{ij}^{(0)} + \ldots \, ,
\eea
where $a^{-1}v^{(1)i}$ are the spatial components of $u^{\mu}$ to leading order and $v_{i}^{(1)} \equiv \delta_{ij} v^{(1)j}$.

In standard cosmological perturbation theory, all perturbations to the metric and matter fields are taken to have the same order-of-smallness, $\epsilon$, such that
\be \label{epsilon}
\epsilon \sim  v^{(1)} \sim g_{\mu \nu}^{(1)} \sim L_C^2 \rho^{(1)} \sim L_C^2 p^{(1)}  \, .
\ee
The reader may note that we have included factors of $L_C^2$ above, so that each of the quantities being compared is dimensionless. This is necessary, strictly speaking, in order to establish that quantities are of the same order of smallness. These additional factors are usually excluded in the literature, but will be important for much of the work we present in this paper.

Substituting both the perturbed metric and the perturbed energy-momentum tensor into Einstein's equation allows us to solve for each of the components of the metric, once an equation of state is specified for the matter fields. In practise this task can be simplified by performing an invariant decomposition of the metric and the velocity field into scalar, divergenceless vector, and transverse-trace-free tensor components. These three types of perturbation do not interact with each other at first order in perturbations, and so the equations that govern each of them can be solved independently of the other two sectors.

The reader will note, due to the considerations at the beginning of this section, that derivatives only affect the order of smallness of a quantity by adding factors of $L_C^{-1}$. That is,
\be 
\dot{\psi} \sim \vert \nabla \psi \vert \sim \frac{\psi}{L_C} \, ,
\ee
where $\psi$ could be either a background quantity, a gravitational potential, or a quantity associated with the matter fields. This is in contrast to the situation in post-Newtonian gravity, as given in Eq. (\ref{smallderivs}). It can also be noted that we require each of the components of the metric only up to first order in perturbations, in order to consistently write the equations of motion of a time-like particle to first order. This is again a departure from the more complicated situation that arises in post-Newtonian gravity. For further explanation of cosmological perturbation theory the reader is referred to the review by Malik \& Wands \cite{malik}.

\subsection{A two-parameter perturbative expansion} \label{pert2para}

In the previous sections we considered post-Newtonian and cosmological perturbative expansions separately. In reality, both types of perturbations are expected to be present in any realistic model of the Universe. We therefore want to construct a two-parameter framework that incorporates them both. We will do this by starting with a FLRW geometry, with the same line-element that appears in Eq. (\ref{flrw}), and then perturbing it using the two parameters $\epsilon$ and $\eta$, which we will take to correspond to the orders of smallness in the cosmological and post-Newtonian expansions, respectively. Such a background is quite standard for cosmological perturbation theory, but little used for post-Newtonian gravity \cite{bruni}. Nevertheless, it is entirely compatible with the discussion in Section \ref{sec:pn}, which we kept general ({\it i.e.} time dependent) in order to allow for this possibility. In fact, a small enough region of perturbed FLRW can be shown to be entirely equivalent to perturbed Minkowski space at both Newtonian \cite{cwoa} and post-Newtonian orders \cite{sanghai1}.

To introduce the idea of a two-parameter expansion, let us start by considering a dimensionless function, or tensorial quantity, $\textbf{F}(x^{\mu})$, that exists in a manifold, $\mathcal{M}$. By expanding in both $\epsilon$ and $\eta$, the smallness parameters associated with our two expansions, we can write this function as
\begin{equation} \label{2ParaPert}
\textbf{F}(x^{\mu}) = \sum_{n,m} \frac{1}{n^{\prime} ! m^{\prime} !}\textbf{F}^{(n,m)}(x^{\mu}) 
\, ,
\end{equation}
where $\textbf{F}^{(n,m)}(x^{\mu})$ are a set of functions that exist in a second manifold $\bar{\mathcal{M}}$, which is diffeomorphic to $\mathcal{M}$. The superscripts $n$ and $m$ on these quantities label their order-of-smallness in $\epsilon$ and $\eta$, respectively. The quantities $n^{\prime}$ and $m^{\prime}$, on the other hand, are set by whether the term in question is leading-order in $\epsilon$ or $\eta$, or next-to-leading-order, {\it etc}. Of course, such an expansion is only possible if both $\epsilon$ and $\eta \ll 1$. Expansions of this kind have already been considered in the literature \cite{bruni2para, bruniNpara}. 

\begin{figure}[t] 
\centering    
    \includegraphics[trim=0cm 21cm 13cm 0cm, clip=true, width=0.4\textwidth]{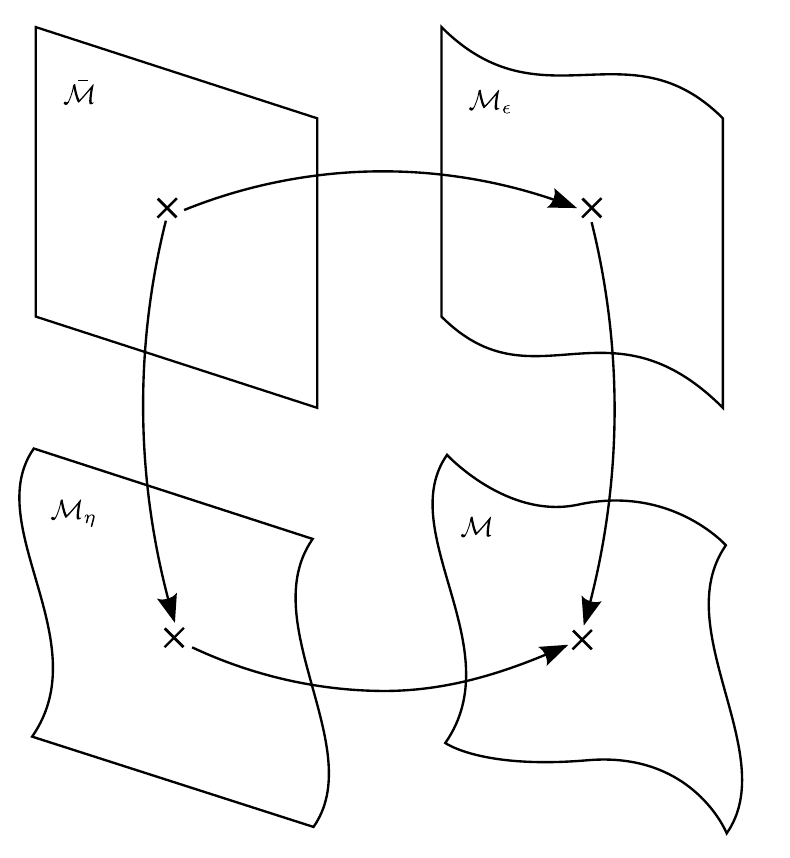}
    \caption{An illustration of the maps between the background manifold $\bar{\mathcal{M}}$, and the manifold of the perturbed space-time, $\mathcal{M}$. The manifolds $\mathcal{M}_{\epsilon}$ and $\mathcal{M}_{\eta}$ correspond to perturbations in $\epsilon$ and $\eta$ only. The two different routes between points on $\bar{\mathcal{M}}$ and $\mathcal{M}$ must be identical if the overall map is invertible.} \label{manifold}
\end{figure}

The geometry of this set-up is illustrated in Fig. \ref{manifold}. The reader should note that perturbed tensors, such as $\mathbf{F}^{(n,m)}$, are pulled-back to the background manifold, $\bar{\mathcal{M}}$, and can therefore be written in terms of the background coordinates, $x^{\mu}$. This then enables us to compare perturbed tensors with unperturbed tensors, just as in single-parameter perturbation theories. Physically, $\textbf{F}(x^{\mu})$ corresponds to a quantity that is close to $\textbf{F}^{(0,0)}(x^{\mu})$, but perturbed in \textit{two} different ways. This is the picture we have in mind when we perturb both the FLRW metric, and the matter fields.

As a simple illustrative example of the scenario we envisage, we could consider a one-dimensional function $\textbf{F}(x)$ that satisfies a given differential equation. If we imagine that $\textbf{F}(x)$ is close to being a sinusoidal wave, then we could write $\mathbf{F}^{(0,0)}(x)=\sin (2 \pi x/\lambda)$. However, if $\textbf{F}(x)$ is not exactly sinusoidal then we may want to calculate the corrections that are required in order to accurately model this function. One way of doing this would be to transform these corrections into a Fourier series, and to split the Fourier modes into those that have a wavelength shorter than $\lambda$, and those that have a wavelength greater than $\lambda$. We can then associate the smallness of the former of these fluctuations with $\eta$, and the latter with $\epsilon$. As long as both $\eta$ and $\epsilon$ are small, we can then use perturbation theory in order to determine the coefficients $\textbf{F}^{(n,m)}$, order by order in smallness. The benefit of using two parameters in this situation is that we are able consider scenarios in which the small-scale corrections behave differently to those that occur on large scales, as happens in cosmology. It also allows us to investigate the way in which small-scale perturbations affect their large-scale counterparts, and {\it vice versa}.

Let us now return to considering cosmology, and continue by expanding both the metric and the matter fields in terms of both $\epsilon$ and $\eta$. These two parameters need not necessarily be of the same size, and, for now, we will keep our expansion general by not assuming anything about the relationship between them. This means, specifically, that we will not assume a relationship of the form $\epsilon = \epsilon (\eta)$, and we will not assume anything about the relationship between the scales $L_N$ and $L_C$ (later on we will restrict ourselves to particular situations of more direct physical interest, in order to write down the field equations, and perform calculations, in a sensible way).

\begin{figure*}
   \includegraphics[trim=0.5cm 13.5cm 0.6cm 0.7cm, clip=true, width=0.9\textwidth]{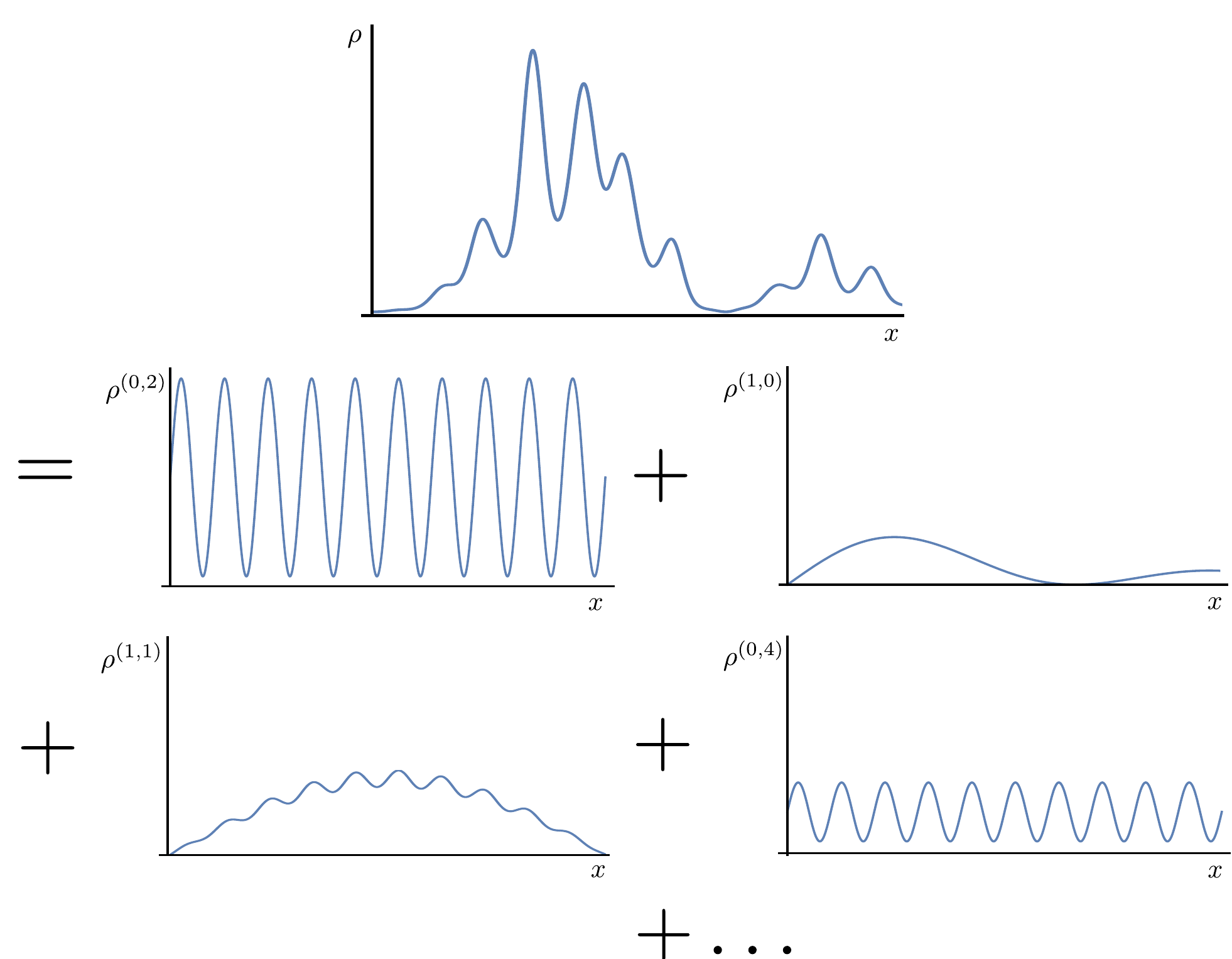}
  \caption{A sketch of the different contributions to the total energy density (top). These contributions include the rest mass energy density (middle left), first cosmological perturbations (middle right), first mixed perturbation (bottom left), and higher-order contributions to internal energy density (bottom right). Smaller contributions to the energy density, at higher-order in perturbation theory, are denoted by the ellipsis. } \label{figendens}
\end{figure*}

Let us start by expanding the energy-momentum tensor, given in Eq. (\ref{e:Tmunu}), in both $\epsilon$ and $\eta$ using Eq. (\ref{2ParaPert}). This gives
\bea \label{e:rhoexp}
\rho  
=  \rho^{(0,2)}+ \rho^{(1,0)}+\rho^{(1,1)} + \rho^{(1,2)}  + {\textstyle\frac{1}{2}}\rho^{(0,4)} + \ldots,
\eea
where
\begin{eqnarray}
\rho^{(n,0)} \sim \frac{\epsilon^n}{L_C^2} \, , \; \rho^{(0,m)} \sim \frac{\eta^m}{L_N^2} \;\; \, \mathrm{and} \;\; \,  \rho^{(n,m)} \sim \frac{\epsilon^n\eta^m}{L_N^2} \, ,
\end{eqnarray}
are the cosmological, post-Newtonian and mixed perturbations of the energy density, respectively. The quantities $\rho^{(0,2)}$ and $\rho^{(0,4)}$ correspond to the energy density in the rest mass of the matter fields and their internal energy density, respectively. Meanwhile, $\rho^{(1,0)}$ is a large-scale cosmological fluctuation in the energy density, and both $\rho^{(1,1)}$ and $\rho^{(1,2)}$ are small-scale perturbations on top of a large-scale fluctuation (or {\it vice versa}). In Fig. \ref{figendens} some of these different contributions to the perturbed energy density are represented visually. 

The reader may note that we have omitted a time-dependent background-level contribution to the energy density, which would otherwise have occurred as $\rho^{(0,0)}(t) \sim L_C^{-2}$. This is intentional, and indeed necessary, if we are to construct a sensible two-parameter expansion in both $\epsilon$ and $\eta$. The reason for this is that such a term, while being usual in single-parameter cosmological perturbation theory, would be highly unusual in post-Newtonian gravity. It would correspond to a contribution to the energy density that is much larger than the rest mass of the matter fields within the space-time. We therefore set $\rho^{(0,0)}=0$, and find out that it is instead the spatial average of $\rho^{(0,2)}$ that plays the role of (what would otherwise be) the background energy density in the Friedmann equations. This will be explained in more detail in Section \ref{SmallAndLarge}.

We derived the expansion of the energy density, given in Eq. (\ref{e:rhoexp}), so that it contains the minimum number of perturbations necessary to describe a two-parameter system. To do this we wrote an initial ansatz for the perturbed energy density that was given by the sum of the post-Newtonian perturbed energy density, the cosmological inhomogeneous perturbed energy density and mixed order perturbations which are products of the leading-order Newtonian and cosmological perturbations. However, after gauge transformation\footnote{After gauge transforming our initial ansatz stress-energy tensor, via the transformations given in Section \ref{Gchoice}, we produced a source of energy density of $\mathcal{O}(\epsilon \eta L_N^{-2})$, see Eq. (\ref{T1100trans}) in that section. This source is of this order because we chose $L_N \sim \eta L_C$. For other relationships between the two length scales there should not be a term $\rho^{(1,1)}$ of $\mathcal{O}(\epsilon \eta L_N^{-2})$ in the expansion of the energy density.} we generated a source of energy density of $\mathcal{O}(\epsilon \eta L_N^{-2})$ . Therefore, we also include a source of energy density order $\rho^{(1,1)}$. This gives the perturbed energy density in Eq. (\ref{e:rhoexp}). This perturbed energy density after gauge transformation is consistent with original energy density, and therefore has the minimal number of perturbations necessary to describe a two-parameter system.

The remaining contributions to the energy-momentum tensor come from the isotropic pressure, $p$, and the peculiar velocity, $v^i$. These are expanded in $\epsilon$ and $\eta$ such that they are the sum of the velocities and pressures used in post-Newtonian gravity and cosmological perturbation theory. No other perturbations are necessary up to the order we wish to consider. Therefore we write
\begin{equation} \label{vexp}
v^i = v^{(0,1)i} + v^{(1,0)i} + \ldots \, ,
\end{equation}
and
\begin{equation} \label{pexp}
p  =  p^{(1,0)} + p^{(0,4)} + \ldots \, , 
\end{equation}
where the peculiar velocity, defined as the spatial part of of the four-velocity $u^{\mu}$ (given in Section \ref{intro}) corresponds to the deviation of the paths of matter fields from the background Hubble flow. If it is zero, then the matter moves only with the expansion of the Universe. If $\eta > \epsilon$ the post-Newtonian velocity $v^{(0,1)i}$ is greater than the velocity allowed by cosmological perturbation theory alone, $v^{(1,0)i}$ (this is the case for the field equations we derive in the following sections).

There are a couple of points that the reader may want to note about these expansions. Firstly, the usual velocity in post-Newtonian gravity does not exactly correspond to the small-scale peculiar velocity $v^{(0,1)i}$. In fact, it is the sum of the small-scale peculiar velocity $v^{(0,1)i}$ \textit{and} the Hubble flow. This is because velocities in normal post-Newtonian gravity are relative to a Minkowski background, whereas in our formalism velocities are peculiar velocities relative to an expanding FLRW space-time. This is an important difference. Secondly, we have not included a contribution to the pressure of the form $p^{(0,2)} \sim \eta^2L_N^{-2}$. Although such a term can be included \cite{sanghai2}, on small scales it corresponds to a barotropic fluid with an energy density comparable to that of dark matter and baryonic matter. While such a fluid could be used to model the effects of radiation in the early Universe, we have chosen to neglect it, in order to model the simpler case of the dust-dominated stages of the Universe's evolution. We instead allow for some small cosmological and post-Newtonian pressure, $p^{(1,0)}$ and $p^{(0,4)}$, respectively.

Let us now consider what happens when derivatives act on the perturbed quantities defined above. We start with the presumption that the rate at which an object changes in space and time can be determined from its order of smallness in $\epsilon$ and $\eta$. If an object is perturbed in $\eta$ only, we will say that it is post-Newtonian. We denote all such objects by $N$, so that $N \sim \eta^m$. Similarly, all objects perturbed in $\epsilon$ only will be called cosmological, and are denoted by $C \sim \epsilon^n$. The remaining objects, perturbed in both $\epsilon$ and $\eta$, will be called mixed, and are denoted by $M \sim \epsilon^n \eta^m$.

Following the discussion in Section \ref{sec:pn}, we will assume that derivatives act on all Newtonian quantities such that
\begin{equation}
 N_{,i} \sim \frac{N}{L_N} \quad \mathrm{and} \quad \dot{N} \sim \frac{\eta N}{L_N} \, .
\end{equation}
Similarly, following the discussion in Section \ref{pertCPT}, we take derivatives to act on all cosmological quantities (includes the scale factor $a(t)$) such that
\begin{equation}
C_{,i} \sim \frac{C}{L_C} \quad \mathrm{and} \quad \dot{C} \sim \frac{C}{L_C} \, .
\end{equation}
It now remains to decide the order of smallness of the derivatives of mixed terms. This is more complicated.

We start our consideration of the derivatives of mixed terms by noting that they vary in space and time on both Newtonian and cosmological length scales, as illustrated in Fig. \ref{figendens}. In order to determine which of these contributions dominate the derivative on a mixed-order quantity we need to relate $L_N$ and $L_C$. In order to do this it is useful to define a new quantity, $l$, such that 
\begin{equation} \label{x}
l \equiv  \frac{L_N}{L_C} \, .
\end{equation}
Also, we observe that we want to consider post-Newtonian perturbed structure, on scales $L_N$, such that the post-Newtonian expansion (around Minkowski space) still holds. For this to be true we need the velocity due to the Hubble flow, $H L_N$, to be smaller than or equal to the peculiar velocities of the constituent objects, $\eta$, hence $H L_N \leqslant \eta$. Otherwise, such systems would have velocities larger than $\eta$ with respect to a Minkowski background, and so post-Newtonian gravity would break down. Given that $H \sim L_C^{-1}$, and using the definition from Eq. (\ref{x}), we then have the requirement
\begin{equation} \label{xleqeta}
l \leqslant \eta \, . 
\end{equation} 
This implies two things: (i) spatial derivatives acting on cosmological terms are strictly smaller than spatial derivatives acting on Newtonian terms, and (ii) time derivatives acting on cosmological terms are strictly less than or equal to time derivative acting on Newtonian terms. Therefore, post-Newtonian spatial and temporal derivatives dominate over or are equal to cosmological ones. Hence we can write 
\begin{equation}
M_{,i} \sim \frac{M}{L_N}\quad \mathrm{and} \quad \dot{M} \sim \frac{\eta M}{L_N} \, ,
\end{equation} 
because, at most, derivatives of mix-ordered terms go like derivatives of post-Newtonian perturbed quantities.

At this point we can make two more comments related to Eq. (\ref{x}) and (\ref{xleqeta}). The first arises because we can write
\bea
 \rho^{(1,0)} \sim  \frac{\epsilon}{L_C^2} \sim \frac{\epsilon l^2}{L_N^2} \, . \label{left}
\eea
This, together with Eq. (\ref{xleqeta}), means that $\rho^{(1,0)} \ll \rho^{(0,2)}$. In other words, the total energy density is \textit{always} dominated by the rest mass of the matter fields on small scales, independent of the relative magnitude of the gravitational potentials on small and large scales. This will be important when it comes to writing the field equations order by order.

The second point is that the above book-keeping of derivatives on Newtonian, cosmological and mixed order terms can be considered in units of either $L_N$ or $L_C$. If we consider the field equations in units of $L_N$ then we relegate certain terms to higher orders, by adding orders of smallness in $\eta$ and $l$. If we consider the field equations in units of $L_C$ we move terms to lower orders, by adding largeness via $\eta^{-1}$ and $l^{-1}$. Either is perfectly acceptable, but we choose to employ the former. This is because it is easier to omit terms which become higher order under a derivative, rather than to go through all possible higher-order terms in order to see which terms might be larger under a derivative.

To complete the description, let us now expand the metric in both $\epsilon$ and $\eta$. Given a background geometry, $g^{(0,0)}_{\mu \nu}$, our two parameter perturbed metric is
\bea
g_{00} &=& g_{00}^{(0,0)} + g_{00}^{(0,2)}  + g_{00}^{(1,0)} + g_{00}^{(1,1)}+ g_{00}^{(1,2)} \nonumber \\ 
&&+ {\textstyle\frac{1}{2}}g_{00}^{(0,4)} + \ldots  \label{g00}\\
&=& -1 + h_{00}^{(0,2)}  + h_{00}^{(1,0)} + h_{00}^{(1,1)}+ h_{00}^{(1,2)} \nonumber \\
&& + {\textstyle\frac{1}{2}}h_{00}^{(0,4)} + \ldots \nonumber  \\[5pt]
g_{ij} &=& g_{ij}^{(0,0)} + g_{ij}^{(0,2)}  + g_{ij}^{(1,0)}+ g_{ij}^{(1,1)} + g_{ij}^{(1,2)} \nonumber \\
&& + {\textstyle\frac{1}{2}}g_{ij}^{(0,4)} + \ldots \label{gij}\\
&=& a^2 \left( \delta_{ij} + h_{ij}^{(0,2)}  +h_{ij}^{(1,0)} + h_{ij}^{(1,1)} + h_{ij}^{(1,2)}    +{\textstyle\frac{1}{2}}h_{ij}^{(0,4)} \right) \nonumber \\ 
&& + \ldots \nonumber \\ [5pt]
g_{0i} &=& g_{0i}^{(1,0)}  + g_{0i}^{(0,3)} + g_{0i}^{(1,2)}+ \ldots  \label{g0i} \\
&=& a \left( h_{0i}^{(1,0)}  + h_{0i}^{(0,3)}+ h_{0i}^{(1,2)}  \right)+ \ldots \, ,  \nonumber
\eea
where in the second line of each of these equations we have specialized to the flat FLRW background, and simultaneously defined the quantities $h_{\mu \nu}$. The orders of magnitude of each of the components of this metric are derived using the method outlined above for post-Newtonian gravity. That is, they are derived from the orders-of-smallness of each of the components of the energy-momentum tensor, together with the orders-of-smallness of space and time derivatives acting on each of the different types of quantities.

We derived the two-parameter expansion of the metric in the same way as the energy density, discussed early in this section, such that the metric contains the minimum number of perturbations necessary to describe a two-parameter system. As with the perturbed energy density, we wrote an initial ansatz for the perturbed metric given by the sum of the FLRW metric, the usual post-Newtonian metric, the cosmologically perturbed metric and mixed order perturbations which are products of the leading order Newtonian and cosmological perturbations. However, after a gauge transformation we produced metric potentials in the $00$, $0i$ and $ij$ parts of the metric at $\mathcal{O}(\epsilon \eta)$, $\mathcal{O}(\epsilon \eta^2)$ and $\mathcal{O}(\epsilon \eta)$, respectively. Therefore, we include metric potentials of order $g^{(1,1)}_{00}$, $g^{(1,1)}_{ij}$ and $g^{(1,2)}_{0i}$ in our new ansatz, giving the perturbed metric above\footnote{The transformation of our initial metric ansatz, via the transformations in Section \ref{Gchoice}, produced metric potentials in the $00$, $ij$ and $0i$ parts of the metric at $\mathcal{O}(\epsilon \eta)$, $\mathcal{O}(\epsilon \eta)$ and $\mathcal{O}(\epsilon \eta^2)$, from Eqs. (\ref{h0011trans}), (\ref{hij11trans}) and (\ref{h0i12trans}), respectively. Again, this was under the choice $l \sim \eta$. Note that for other relationships between the two length scales $L_N$ and $L_C$ there should not be terms $g^{(1,1)}_{00}$ and $g^{(1,1)}_{ij}$ at order $\mathcal{O}(\epsilon \eta)$. However, for all relationships between $L_N$ and $L_C$ there would exist a metric potential at order $g^{(1,2)}_{0i}$, after gauge transformation.}. Now, the new perturbed metric after gauge transformation is consistent with original metric, and therefore has the minimal number of perturbations necessary to describe a two-parameter system.

The full expressions for the perturbed energy-momentum and Ricci tensors are given in the Appendix, and will be used in Section \ref{sec:fe}. 

\section{Observational justification} \label{obsjust}

In the previous section we considered the different ways that perturbation theory can be applied to gravitational fields on both horizon-sized and sub-horizon-sized regions of space-time. This resulted in a derivation of both the post-Newtonian and cosmological perturbation theories, using little more than the fact that Einstein's equations can be written as null wave equations. We then considered how these two different expansions could be formally combined into a two-parameter expansion that could be used to describe the Universe on both large and small scales. Throughout all of this we tried to keep the discussion as general as possible, without specifying any specific relationship between either the expansion parameters $\epsilon$ and $\eta$, or the length scales $L_C$ and $L_N$.

In this section we consider observations of the specific astrophysical systems that exist on different scales in the Universe. The aim of this is to see which types of systems are best described by post-Newtonian expansions, and which are best described using cosmological perturbation theory. This allows us to consider the physical scenarios that could potentially be described using our two-parameter expansion, as well as the particular values of $\epsilon$ and $\eta$ that are appropriate in each case. Of course, each pair of systems also comes with its own values of $L_C$ and $L_N$, which can also be related to the expansion parameters. Once we have all of this information at hand, we can then write down the field equations of our two-parameter expansion, order by order in the appropriate parameters.

\subsection{Post-Newtonian gravity }

Post-Newtonian perturbative expansions are usually applied to describe the gravitational physics of astrophysical bodies that range in size from binary pulsar systems (about a million kilometres), to the size of the orbits of the planets in our solar system (a few hundred million kilometres). Let us begin by considering these systems, before moving on to the larger astrophysical systems that are of more interest for cosmology. To do this, we will quote estimates for the largest velocities that occur within them, and compare these to estimates of the largest gravitational potentials that we can find using the order-of-magnitude estimator
\be \label{Newtonslaw}
U = \frac{GM_N}{c^2L_N } \, ,
\ee
where $M_N$ and $L_N$ are observational estimates of the mass and length scale of the system, and are in units of kilograms and meters, respectively. This will allow us to estimate $\eta$, as well as establish whether or not a given system is indeed suitably described using a post-Newtonian perturbative expansion. The results are summarized in Table \ref{table}.

The largest velocities in the Solar System correspond to coronal mass ejections, which can erupt at up to $450 \mathrm{km\, s}^{-1}$ (see p. 375 of \cite{allen}). This corresponds to $v \sim 10^{-3}$, in units where $c=1$. As well as this, the mass of the Sun is about $M_{\odot} \sim 2 \times 10^{30} \mathrm{kg}$, and its radius is approximately $L_N \sim L_{\odot}\sim 7 \times 10^{8} \mathrm{m}$. This means that Eq. (\ref{Newtonslaw}) implies $U \sim 10^{-6}$. This means that the post-Newtonian expansion is indeed applicable, because $v^2 \sim U$, as expected from Eq. (\ref{Usim2}). It also means that the value of the expansion parameter in this system is given by $\eta \sim 10^{-3}$.

There are a number of systems that one could consider above the scale of the Sun, but to speed the discussion let us move directly up to the scale of spiral galaxies. These systems are typically made up of billions of stars, and typically have a bulge, a disk, and a dark matter halo. The observed velocities of stars can be as high as $300\mathrm{km\, s}^{-1}$ (see p. 571, 578 \& 580 of \cite{allen}). This again corresponds to $v \sim 10^{-3}$. If we consider a bulge of radius $L_N \sim 10\mathrm{kpc}$, and mass $M_N \sim 10^{11} M_{\odot}$, then this gives $U \sim 10^{-6}$. We again have $v^2 \sim U$, meaning that a post-Newtonian perturbative expansion seems appropriate to describe the gravitational field, and we again have $\eta \sim 10^{-3}$.

Typical galaxy groups contain $3$-$30$ galaxies that are gravitationally bound, and it is estimated that $\sim 55\%$ of galaxies exist within groups. The maximum radial dispersion in groups of galaxies is observed to be about $500\mathrm{km\, s}^{-1}$ (see p. 614 of \cite{allen}), again implying $v \sim 10^{-3}$. We estimate the mass of a typical group, including dark matter, is $M_N \sim 10^{13} M_{\odot}$, and that the radius of a typical group is $L_N \sim 0.8 \mathrm{Mpc}$ (this is an average of the range given in p. 614 \cite{allen}). This implies that $U \sim 10^{-6}$ in galaxy groups, and that the post-Newtonian perturbative expansion seems to apply here as well. We even have $\eta \sim 10^{-3}$, as above.

\begin{table}[t]
\centering
\label{table}
\begin{tabular}{l|llll}
    \hline
    System & $\;\;v$  & $L_N/\mathrm{Mpc}$& $M_N/M_{\odot}\;$ &  $U$    \\[5pt] 
    \hline
    Sun & $\;\;10^{-3}\quad$  & $2 \times 10^{-14} \quad$   & $ 1\quad$  & $10^{-6}\quad$ \\[5pt]
    Galaxy & $\;\;10^{-3}$ & $10^{-2} $   & $10^{12} $  & $10^{-6}$ \\[5pt]
    Group & $\;\;10^{-3}$ & $0.8 $   & $10^{13} $ & $10^{-6}$ \\[5pt]
    Cluster  & $\;\;10^{-2.5}$ & $2 $   & $10^{15} $ & $10^{-5}$ \\[5pt]
    Supercluster$\quad$ & $\;\;10^{-2.5}$ & $100 $   & $10^{16}$ & $10^{-5}$ \\[5pt]
    \hline  
\end{tabular}
\caption{Summary of the magnitude of $v$ and $U$ in a variety of gravitational bound systems, covering a wide range of different scales.} \label{table}
\end{table}

Moving up in scale still further, we have clusters of galaxies. Typical galaxy clusters contain $30$-$300$ gravitationally bound galaxies. The dispersion velocities of galaxies within clusters can be as large as $1400\mathrm{km\, s}^{-1}$, or $v \sim 10^{-2.5}$ in units where $c=1$. We take the mass of a typical cluster to be about $M_N \sim 10^{15} M_{\odot}$, and the average radius to be around $L_N \sim 2 \mathrm{Mpc}$ (averages of quantities given on p. 614 of \cite{allen}). Similarly we average to find the typical radius of a cluster which is around $L_N \sim L_{\mathrm{cluster}} \sim 2\mathrm{Mpc}$. The maximum gravitational potentials expected in clusters are therefore $U \sim 10^{-5}$. We again have $v^2 \sim U$, but now with $\eta \sim 10^{-2.5}$.

Super-clusters are the largest virialized objects we currently observe in the Universe. They make up the filaments and walls that form the cosmic web, and are made from clusters, groups and other smaller gravitationally bound systems. Observations show that peculiar velocities within of our own local supercluster are around $1000 \mathrm{km\, s}^{-1}$ \cite{nature,nature2}, which corresponds to $v \sim 10^{-2.5}$. There are typically $2$-$15$ clusters per supercluster, which implies the mass of the supercluster is at least $10^{16} M_{\odot}$ (see p. 635 of \cite{allen}). They have typical scales of $L_N \sim 100 \mathrm{Mpc}$. This gives $U \sim 10^{-5}$. Even on these extraordinarily large scales, we have $v^2 \sim U$ and $\eta \sim 10^{-2.5}$.

It is interesting to note the maximum amplitude of the gravitational potential is roughly $\sim 10^{-5}$ for all of the systems considered above. This ranges over just about all astrophysical objects, from the Sun to our local supercluster. We therefore have an expansion parameter $\eta \sim 10^{-3}$ for all of these systems. The similarity in the size of the gravitational potential, no matter what system is being considered, indicates that the mass of the system under consideration increases approximately in proportion to its length scale. This type of self-similarity will break down whenever a system's mass is much larger than about $10^{-5}$ of its length scale, at which point we expect the post-Newtonian expansion should start to break down. This happens, for example, in the case of neutron stars.

Although post-Newtonian perturbation theory appears to be applicable to superclusters, we do not expect it to be valid on scales that are much larger. This is because the square of the velocity due to the Hubble flow starts to become comparable to the order of the Newtonian potentials, {\it i.e.} $H ^2 L_N^2 \sim 10^{-5}$. Going to even larger scales would therefore mean that the square of the Hubble flow velocity would start to exceed the magnitude of the gravitational potentials. If this is the case then post-Newtonian expansions are no longer applicable, and cosmological perturbation theory must be used. It is expected that the next generation of surveys, such as Euclid, LSST and SKA, will start to probe this new regime.

\subsection{Cosmological perturbation theory}

Let us now consider the largest of all scales in the observable Universe; those comparable to the size of the horizon. In terms of the CMB, this corresponds to about $1$ degree. In the late Universe it corresponds to scales around $30$Gpc. In this case we expect the cosmological perturbation theory expansion outlined in Section \ref{pertCPT} to be applicable. The principle distinction between the size of the perturbed quantities in this expansion, when compared to the post-Newtonian expansion, is that time derivatives do not add any extra orders of smallness. This means that velocity cannot be used as an expansion parameter. The separation of objects is instead dominated by the Hubble flow, with only small peculiar velocities (of the order of gravitational potentials) being allowed in addition.

The discussion of superclusters, in the previous section, should already have made it clear that cosmological perturbation theory is not the appropriate framework for discussing the dynamics of astrophysical systems that exist below $\sim 100$Mpc. This is essentially because the time variation of both gravitational and matter fields are slow compared to their variation in space, meaning that $U \sim v^2$. On larger scales, however, we expect to find $U \sim v$. There do not currently exist any galaxy surveys that probe these scales directly, but we can use the CMB to justify the application of cosmological perturbation theory on horizon-sized length scales and above.

The temperature fluctuations in the CMB, after the dipole has been subtracted, are all at the level of about $10^{-5}$ \cite{COBE}. The main contribution to these fluctuations, on large scales, is expected to come from the Sachs-Wolfe effect. This is essentially a redshifting of the CMB radiation as it escapes the gravitational potentials that existed at the surface of last scattering, and the redshift is of course related to the temperature in a well-known way. We therefore expect
\be
\frac{\delta T}{T} \sim U \, ,
\ee
where $U$ should be understood as a typical gravitational potential at last scattering. The observations of the temperature fluctuations at the level of $1$ part in $10^5$ therefore very directly imply that gravitational potentials at last scattering were of the size $U \sim 10^{-5}$.

If we now consider the polarization of the CMB, then we can gain information about the magnitude of peculiar velocities at last scattering. This is because polarization of the CMB radiation, $\mathcal{E}$, is primarily due to quadropole anisotropy in the velocity field of the plasma at last scattering \cite{rees}. We expect the mean-free path of photons at last scattering to be of the order of the inverse Hubble rate (so that $1/n_e \sigma_t \sim L_C$, where $n_e$ is the number density of electrons, and $\sigma_t$ is the Thomson cross section). The polarisation is therefore given by
\be
\mathcal{E} \sim \Delta v \, ,
\ee
where $\Delta v$ is the difference in peculiar velocity of matter, in orthogonal directions on the sky (for details see \cite{rees}). Observations of CMB polarization now measure $\mathcal{E} \sim 10^{-6}$ \cite{polarisation}, which means that peculiar velocities at last scattering are of order $v \sim 10^{-6}$.

Taken together, these observations therefore suggest that $v \sim U$ on horizon-sized scales, as expected. These results clearly indicate that a post-Newtonian expansion is {\it not} the appropriate framework to be describing gravity on these scales, and that cosmological perturbation theory should be used instead. What is more, it can be seen that the expansion parameter for the cosmological perturbation theory should be of magnitude $\epsilon \sim 10^{-5}$. Although it has not yet been directly observed, we very strongly expect similar results to hold at and above $\sim 1$Gpc in the late Universe.

\subsection{A realistic universe}

In the preceding sections, we found that planetary systems, galaxies, groups, clusters and superclusters are all well described by post-Newtonian gravity. That is, their observed velocities and inferred gravitational potentials satisfy $v^2 \sim U \sim 10^{-5}$. Additionally, we find that observed fluctuations on the scale of the horizon are well described by cosmological perturbation theory, as $v\sim U \sim 10^{-5}$. This very strongly indicates that post-Newtonian gravity cannot be used to describe structure on the scale of the horizon, and that cosmological perturbation theory cannot be used to describe non-linear structure on the scale of $100$Mpc or less.

In order to model a realistic Universe, that has non-linear structure on small scales, as well as linear structure on large scales, we therefore need to expand in both $\epsilon$ and $\eta$. This is exactly the type of two parameter expansion that we wish to formulate in this paper. In what follows, we will take $\epsilon \sim \eta^2 \sim 10^{-5}$, as this seems to fit almost all large astrophysical structures that exist in the Universe, and that we wish to describe with our formalism. We will also take $L_C\sim 30$Gpc and $L_N \sim 100$Mpc, so $l \sim \eta$. These length scales correspond to the horizon size at the present time, and the saturation of the bound in Eq. (\ref{xleqeta}). This latter length scale also happens to roughly correspond to that of the largest gravitationally bound objects that have so far been observed to exist in the Universe. For this system, in what follows, we will write the field equations order by order in a two parameter expansion.

\section{Field equations}
\label{sec:fe}

It is straightforward to expand the field equations (\ref{EE}) in both $\epsilon$ and $\eta$, but the results are somewhat lengthy. This is partly due to the fact that we are using two parameters in our perturbative expansion, but is also a result of the freedom in choosing coordinates that exists within General Relativity. Nevertheless, we want to present our results in the most general form possible. We therefore write out the full versions of the Ricci tensor and energy-momentum tensor in the Appendix, where these objects are perturbed in both $\epsilon$ and $\eta$. The form of these equations is particularly complicated not only because each component of every tensor contains a large number of terms, but because each term is itself associated with a different length scale (or set of scales).

In practise, we want to apply our formalism to specific examples of physical interest. Once such an example scenario has been chosen, then the expansion parameters and length scales can be written in terms of one another. This reduces the complexity, and allows the field equations to be written out explicitly, and without ambiguity. In this section we will present results for the choice
\be
\epsilon \sim \eta^2 \sim \frac{L_N^2}{L_C^2} \sim 10^{-5} \, , 
\ee
as described at the end of Section \ref{obsjust}. These results will be presented without fixing coordinates to any particular gauge, and are therefore still quite lengthy. In Section \ref{Gchoice} we will exploit the gauge freedom associated with coordinate re-parameterization, and use this to present the same field equations in a much more compact form in Section \ref{sec:gaugeinvariants}.

At this stage it is useful to define some new notation, so that we can present the trace-free part of various quantities in the most efficient way possible. We define angular brackets on a pair of indices to mean that they are symmetric and trace-free, such that
\be
\label{tfdef}
\mathcal{T}_{\langle ij \rangle} \equiv \mathcal{T}_{(ij)} - \frac{1}{3} \delta_{ij} \mathcal{T}_{kk} \, ,
\ee
where $\mathcal{T}$ is a rank-2 tensor, and where indices are now being raised and lowered with the Kronecker delta, $\delta_{ij}$. The round brackets in this expression denote symmetrization, and repeated indices are summed over, as usual. We will also use vertical lines around indices if they are to be excluded from a symmetrization or trace-free operation.

Additionally, we define a symmetric and trace-free second derivative operator by the following equation:
\bea
\label{dtfdef}
D_{ij} \varphi \equiv \varphi_{,(ij)} - \frac{1}{3} \delta_{ij} \nabla^2 \varphi \, ,
\eea
where $\varphi$ is any tensorial quantity (not necessarily a scalar), and where $\nabla$ represents the Laplacian on Euclidean space. We will use this notation to write out the trace and trace-free parts of the field equations, order by order in perturbations.

\begin{widetext}

\subsection{Background-order potentials} \label{bkefes}

The leading-order part of the field equations, in our formalism, is not at zeroth order in the expansion parameters. Instead, we find that it comes in at $\mathcal{O}(\eta^2 L_{N}^{-2})$. The leading-order part of the $00$-field equation is therefore given by
\begin{equation} \label{e00N}
\frac{\ddot{a}}{a} + \frac{1}{6 a^2} \nabla^2 h^{(0,2)}_{00} = - \frac{4 \pi}{3}  \rho^{(0,2)} \, .
\end{equation}
This equation results from Eqs. (\ref{R0000}), (\ref{R0002}) and (\ref{T0002}), and is a combination of both the Raychaudhuri equation and the Newton-Poisson equation. It is interesting to see that the rest mass density, $\rho^{(0,2)}$, is the source of both the Newtonian gravitational field and the large-scale acceleration equation. This is compatible with the usual understanding of how these phenomena are generated, but it is not usual to see them occurring in the same equation, at the same order in perturbations.

At the same order of accuracy, we find that the leading-order contribution to the trace of the $ij$-field equations is given by
\bea
\frac{\dot{a}^2}{a^2} - \frac{1}{6 a^2} \left(\nabla^2 h^{(0,2)}_{ii} -h^{(0,2)}_{ij,ij} \right) = \frac{8\pi}{3} \rho^{(0,2)} \, . \label{ijtrace}
\eea
This equation is derived from Eqs. (\ref{Rij00}), (\ref{Rij02}) and (\ref{T0002}), and is a combination of the Friedmann equation and the Newton-Poisson equation for the trace of the post-Newtonian potential $h_{ii}^{(0,2)}$. Again, it is somewhat unusual to see a mixture of what might otherwise be considered background and first-order terms, if one were using single-parameter cosmological perturbation theory.
 
Finally, the trace-free part of the $ij$-field equations is also at $\mathcal{O}(\eta^2 L_{N}^{-2})$, and is given by
\bea
D_{ij} \left( h^{(0,2)}_{00} - h^{(0,2)}_{kk} \right) + 2 h^{(0,2)}_{k\langle i,j\rangle k} - \nabla^2 h^{(0,2)}_{\langle ij \rangle} = 0 \, , \label{ijtracefree}
\eea
where we have made use of the notation introduced in Eqs. (\ref{tfdef}) and (\ref{dtfdef}). This equation looks like the quasi-static limit of a first-order equation from cosmological perturbation theory.

\subsection{Vector potentials} \label{bbkefes}

Now let us consider the $0i$-field equations, which usually result in the governing equations for the vector gravitational potentials. The leading-order contribution to these equations comes in at $\mathcal{O}(\eta^3 L_N^{-2})$, and is given by
\bea \label{e0iPN}
\nabla^2 h^{(0,3)}_{0i} -  h^{(0,3)}_{0j,ij} - a \dot{h}^{(0,2)}_{ij,j} + a \dot{h}^{(0,2)}_{jj,i} + 2\dot{a} h^{(0,2)}_{00,i} = 16 \pi a^2 \rho^{(0,2)}v^{(0,1)}_i \, .
\eea
This equation is the result of using Eqs. (\ref{R0i02}), (\ref{R0i03}) and (\ref{T0i03}), from the Appendix. It can be considered as the governing equation for small-scale vector potentials, which will source phenomena such as the Lense-Thirring effect.

At next-to-leading-order in the $0i$-field equation, at $\mathcal{O}(\eta^4 L_N^{-2})$, we find from Eqs. (\ref{R0i10})-(\ref{R0i12}) and (\ref{T0i12}) that
\bea \label{0i1.5efe}
\hspace{-4pt}
&& \nabla^2 \left(h^{(1,0)}_{0i} +h^{(1,2)}_{0i}\right)- \left(h^{(1,0)}_{0j} + h^{(1,2)}_{0j}\right)_{,ij}  -  h^{(1,0)}_{0j}h^{(0,2)}_{00,ij}   - a \left(h^{(1,0)}_{ij}+h^{(1,1)}_{ij}\right)^{\cdot}_{,j} 
+ a  \left(h^{(1,0)}_{jj}+h^{(1,1)}_{jj}\right)^{\cdot}_{,i} \\
&&+2 \dot{a} \left(h^{(1,0)}_{00}+h^{(1,1)}_{00}\right)_{,i}  - 2 h^{(1,0)}_{0i} \left( 2 \dot{a}^2  + a \ddot{a} \right) \nonumber \\
&=&
8 \pi a^2  \left(2\rho^{(1,1)}v^{(0,1)}_i + \rho^{(0,2)} \left( h^{(1,0)}_{0i} + 2 v^{(1,0)}_i  \right)\right) \, . \nonumber
\eea
This equation can be thought of as the governing expression for the large-scale vector potentials. It is more complicated than Eq. (\ref{e0iPN}), and shows that non-linear gravitational effects could potentially source the growth of large-scale vector potentials at late times.

\subsection{Higher-order scalar potentials}

The next-to-leading-order 00-field equation is $\mathcal{O}( \eta^3 L_N^{-2})$, and is given by
\begin{equation} \label{e00N}
 \frac{1}{6 a^2} \nabla^2 h^{(1,1)}_{00} = - \frac{4 \pi}{3}  \rho^{(1,1)} \, .
\end{equation}
This is a Newton-Poisson equation, derived from Eqs. (\ref{R0011}) and (\ref{T0011}). It is sourced only by a mixed order energy density $\rho^{(1,1)}$. This is not usual because the Newton-Poisson equation is normally only at leading order and, of course, is not normally associated with a mixed-order perturbed quantity.

The metric perturbations that correspond to cosmological scalar potentials are $h^{(1,0)}_{00}$ and $h^{(1,0)}_{ii}$. The governing equations for both of these perturbations occur with post-Newtonian and mixed order potentials at $\mathcal{O}(\eta^4 L_N^{-2})$, just as was the case for the vector potentials considered above, and as expected. From the $00$-field equation, at this order, we therefore find that
\bea \label{e00PN}
&&\nabla^2 \left( h^{(1,0)}_{00} + h^{(1,2)}_{00} + \frac{1}{2}h^{(0,4)}_{00} \right) 
+ \frac{1}{2} \left(\nabla h^{(0,2)}_{00} \right)^2  
+ a^2 \left( h^{(0,2)}_{ii} + h^{(1,0)}_{ii} \right)^{\cdot \cdot}
- 2 \left[ a \left( h^{(0,3)}_{0i}  + h^{(1,0)}_{0i} \right)_{,i} \right]^{\cdot}  
\nonumber \\[5pt] 
&& + 2a \dot{a} \left( h^{(0,2)}_{ii} +h^{(1,0)}_{ii} \right)^{\cdot} 
- \frac{1}{2} h^{(0,2)}_{00,i} \left( 2h^{(0,2)}_{ij,j} -h^{(0,2)}_{jj,i} \right) 
- h^{(0,2)}_{00,ij} \left( h^{(1,0)}_{ij} + h^{(0,2)}_{ij} \right) 
+ 3 a \dot{a} \left( h^{(0,2)}_{00} +h^{(1,0)}_{00} \right)^{\cdot} 
\nonumber \\[5pt] 
&=& -8 \pi a^2  \left[ \rho^{(1,0)}  + \rho^{(1,2)} + \frac{1}{2}\rho^{(0,4)} - \rho^{(0,2)} \left( h^{(1,0)}_{00} + h^{(0,2)}_{00} \right) + 3 \left( p^{(1,0)} + p^{(0,4)} \right) + 2\left( v^{(0,1)}_i \right)^2 \rho^{(0,2)} \right] \, , 
\eea 
which has been derived using Eqs. (\ref{R0003})-(\ref{R0010}), (\ref{R0012}), (\ref{T0004}), (\ref{T0010}), (\ref{T0012}), (\ref{Tij04}) and (\ref{Tij10}) from the Appendix. There are a number of interesting things to note about this equation. These include the fact that the cosmological scalar $h^{(1,0)}_{00}$ is sourced by terms that are quadratic in the small-scale Newtonian potential, $h^{(0,2)}_{00}$, as well as terms that are linear in the vector potential, $h^{(0,3)}_{0i}$, and post-Newtonian potential $h^{(0,4)}_{00}$. This kind of mixing in scales and modes is a product of the approach we have used in our two-parameter perturbative expansion and could explain why studies of second-order gravitational fields in cosmological perturbation theory average to the size of first order gravitational fields \cite{Alan, Chris2, Kolb1, Rasanen, Adamek}. It suggests that interesting relativistic phenomenology could result at linear order on large scales in the late Universe.

The $ij$-field equation, at $\mathcal{O}(\eta^3 L_N^{-2})$, can be split into its trace and trace-free parts. The trace-free part will be presented in the next subsection. The trace gives
\bea
 - \frac{1}{6 a^2} \left(\nabla^2 h^{(1,1)}_{ii} -h^{(1,1)}_{ij,ij} \right) = \frac{8\pi}{3} \rho^{(1,1)} \, . \label{ij11trace}
\eea
This equation is derived from Eqs. (\ref{Rij11}) and (\ref{T0011}) and is a Poisson equation for the trace of the mixed order potential $h^{(1,1)}_{ii}$. Again, this is not usual because such an equation is normally at post-Newtonian order and is normally not associated with a mixed-order quantity.

The $ij$-field equation, at $\mathcal{O}(\eta^4 L_N^{-2})$, can also be split into its trace and trace-free parts. The trace-free part will be presented in the next subsection. The trace gives
\bea \label{eijPNtrace}
&& \left(\delta_{ij} \nabla^2- \partial_i \partial_j\right) \left( h^{(1,0)}_{ij} + h^{(1,2)}_{ij} + \frac{1}{2}h^{(0,4)}_{ij} \right) 
- \left(2\dot{a}^2 +   a \ddot{a}\right) \left( h_{ii}^{(1,0)} +h_{ii}^{(0,2)}+ 3 h^{(1,0)}_{00}+ 3 h^{(0,2)}_{00} \right)  \nonumber \\[5pt] 
&& +4 \dot{a}\left( h^{(1,0)}_{0i} + h^{(0,3)}_{0i} \right)_{,i}  
 -2a\dot{a}\left( h_{ii}^{(1,0)}+ h_{ii}^{(0,2)} \right)\dot{} \nonumber \\[5pt] 
&=& -4 \pi a^2 \left[ 4 \left( \rho^{(1,0)} +  \frac{1}{2}\rho^{(0,4)} + \rho^{(1,2)} \right)
+\rho^{(0,2)} \left( h^{(1,0)}_{ii} + h^{(0,2)}_{ii} - h^{(1,0)}_{00} - h^{(0,2)}_{00} + 4  \left( v^{(0,1)}_i \right)^2 \right) \right] + \mathcal{A} \, , 
\eea
where we have simplified this expression using Eq. (\ref{e00PN}) multiplied by a factor of $a^2$. The $\mathcal{A}$ in Eq. (\ref{eijPNtrace}) represents the sum of all terms that are quadratic in lower-order potentials, and is given by
\bea 
\mathcal{A} 
&\equiv & 
\frac{3}{4} \left( h^{(0,2)}_{ij, k} \right)^2 
+ h^{(0,2)}_{ij, j} \left( h^{(0,2)}_{kk,i} -  h^{(0,2)}_{ik,k} \right) 
- \frac{1}{2}h^{(0,2)}_{ij,k} h^{(0,2)}_{ik,j} 
- \frac{1}{4}h^{(0,2)}_{ii,j}h^{(0,2)}_{kk,j} 
+ \frac{1}{2}\nabla^2 h^{(0,2)}_{00} \left( h^{(1,0)}_{00} + h^{(0,2)}_{00} \right) \nonumber \\[5pt] 
&&+ \frac{1}{2} \left( h^{(0,2)}_{00,ij}+ \nabla^2 h^{(0,2)}_{ij} \right) \left( h^{(1,0)}_{ij} + h^{(0,2)}_{ij} \right) 
+ \left( \frac{1}{2}h^{(0,2)}_{ii,jk} -  h^{(0,2)}_{ij,ik} \right) \left( h^{(0,2)}_{jk} + h^{(1,0)}_{jk} \right)  \, . 
\eea
These expressions result from Eqs. (\ref{Rij03})-(\ref{Rij10}), (\ref{Rij12}), (\ref{T0004}), (\ref{T0010}), (\ref{T0012}), (\ref{Tij04}) and (\ref{Tij10}), in the Appendix. If $\mathcal{A}$ is non-zero, then this indicates that non-linear relativistic effects could be important in the determination of scalar gravitational fields on large scales. One may also note that small-scale peculiar velocities are now a source for linear cosmological scalar gravitational fields.

\subsection{Tensor potentials}

The next-to-leading-order trace-free $ij$-field equation is at $\mathcal{O}(\eta^3L_N^{-2})$ and given by
\bea
D_{ij} \left( h^{(1,1)}_{00} - h^{(1,1)}_{kk} \right) + 2 h^{(1,1)}_{k\langle i,j\rangle k} - \nabla^2 h^{(1,1)}_{\langle ij \rangle} = 0 \, , \label{ij11tracefree}
\eea
where we have used Eqs. (\ref{Rij11}) and (\ref{T0011}). We note that this equation has the same form at the lowest order trace-free $ij$-field equation, given in Eq. (\ref{0i1.5efe}).

The remaining part of the field equations that we wish to consider is the trace-free part of the $ij$-component. At $\mathcal{O}(\eta^4 L_N^{-2})$ we find that this equation is given by
\bea  \label{eijPNtracefree}
&&\;\;
\nabla^2 \left( h^{(1,0)}_{\langle ij \rangle} +h^{(1,2)}_{\langle ij \rangle} + \frac{1}{2}h^{(0,4)}_{\langle ij \rangle} \right) 
- D_{ij} \left( h^{(1,0)}_{00} + h^{(1,2)}_{00} +\frac{1}{2}h^{(0,4)}_{00} - h^{(1,0)}_{kk} - h^{(1,2)}_{kk} - \frac{1}{2}h^{(0,4)}_{kk} \right) 
\nonumber \\[5pt] 
&&\;\;
- 2 \left( h^{(1,0)}_{k\langle i} +h^{(1,2)}_{k\langle i} +\frac{1}{2}h^{(0,4)}_{k\langle i} \right)_{,j \rangle k}   
- a^2 \left( h^{(1,0)}_{\langle ij \rangle} + h^{(0,2)}_{\langle ij \rangle } \right)^{\cdot \cdot}
\nonumber \\[5pt] 
&&\;\;
- 2 \left(2\dot{a}^2+ a\ddot{a}\right) \left( h^{(1,0)}_{\langle ij \rangle } + h^{(0,2)}_{\langle ij \rangle } \right)  
- 3a\dot{a} \left( h^{(1,0)}_{\langle ij \rangle}   + h^{(0,2)}_{\langle ij \rangle } \right)^{\cdot}   
+ \frac{2}{a} \left[ a^2 \left( h^{(1,0)}_{0\langle i} +h^{(0,3)}_{0\langle i} \right)\right]^{\cdot}_{,j \rangle }  
 \nonumber \\[5pt] 
&&= -8 \pi a^2 \rho^{(0,2)} \left[  h^{(1,0)}_{\langle ij \rangle} + h^{(0,2)}_{\langle ij \rangle} + 2 v^{(0,1)}_{\langle i} v^{(0,1)}_{j\rangle} \right] + \mathcal{B}_{ij}  \, , 
\eea
where we used $\mathcal{B}_{ij}$ to denote the summation of all terms that are quadratic in lower-order potentials, such that
\bea \label{Bij}
\mathcal{B}_{ij} &\equiv &  
\frac{1}{2} h^{(0,2)}_{00, \langle i \vert}h^{(0,2)}_{00, \vert j \rangle}  
+\frac{1}{2} h^{(0,2)}_{kl, \langle i\vert} h^{(0,2)}_{kl, \vert j \rangle}  
+  D_{ij}h^{(0,2)}_{00} \left( h^{(1,0)}_{00} + h^{(0,2)}_{00} \right)  
+\frac{1}{2} \left( h^{(0,2)}_{00,k} + 2h^{(0,2)}_{kl,l} -h^{(0,2)}_{ll,k} \right) \left( h^{(0,2)}_{\langle ij \rangle,k} -2h^{(0,2)}_{k\langle i,j\rangle} \right) 
\nonumber \\[5pt] 
&&+ \left( D_{ij}h^{(0,2)}_{kl} + h^{(0,2)}_{\langle ij \rangle,kl} -2h^{(0,2)}_{k\langle i,j \rangle l} \right) \left( h^{(1,0)}_{kl} + h^{(0,2)}_{kl} \right)  
+ h^{(0,2)}_{\langle i \vert k,l} \left( h^{(0,2)}_{\vert j\rangle k,l} -h^{(0,2)}_{\vert j\rangle l,k} \right)  \, . 
\eea
These expressions also result from Eqs. (\ref{Rij03})-(\ref{Rij10}), (\ref{Rij12}), (\ref{T0004}), (\ref{T0010}), (\ref{T0012}), (\ref{Tij04}) and (\ref{Tij10}), in the Appendix. They show that trace-free large-scale tensor potentials are, in this formalism, sourced by peculiar velocities, as well as by terms that are quadratic in lower-order potentials. This again indicates the possibility of mode mixing between scales, and the sourcing of gravitational phenomena in ways that are impossible at first order in standard cosmological perturbation theory.

In the next section we will consider how gauge transformations affect the perturbations that we have been considering. This information will then be used to simplify the field equations that are given above, as well as to present them in a gauge-invariant form.

\section{Infinitesimal coordinate transformations}
\label{Gchoice}

General relativity is a diffeomorphism covariant theory, meaning that the form of the tensor equations that we use to describe it must be valid for any set of coordinates. Now, diffeomorphisms obey a strict group structure, which guarantee that we can transform any given solution into a new set of coordinates, and that the result will still obey Einstein's equations. When considering general perturbations about a fixed background, this freedom in coordinate re-parameterization is referred to as a ``gauge freedom''.

When it comes to solving Einstein's equations (\ref{EE}), coordinate re-parameterization invariance, and gauge freedom, are both a blessing and a curse. In general, they mean that perturbations, such as perturbations to the metric, contain not only the essential degrees of freedom required to describe the physical situation at hand, but also a number of superfluous degrees of freedom that relate only to the coordinates used to describe the problem. However, while it takes some care to remove these extra degrees of freedom, the process of doing so can be used to simplify the equations that result. This is especially welcome in our case, as the equations presented in Section \ref{sec:fe} are particularly unwieldy.

In this section we will outline the how gauge transformations should be performed in our two-parameter perturbative expansion. The form of these transformations will then be used in Section \ref{sec:gaugeinvariants} to construct a set of variables that have the superfluous gauge freedoms removed. This will allow us not only to write the field equations in a more compact form, but also to present a set of equations that represents only the degrees of freedom required to characterise the physical problem itself. Additionally, a full understanding of the gauge transformations of the matter and metric fields also allows us to identify the terms that should appear in Eqs. (\ref{e:rhoexp}), (\ref{vexp}), (\ref{pexp}), (\ref{g00})-(\ref{g0i}). 

\subsection{Mathematical structure of gauge transformations}

The general form of an infinitesimal gauge transformation can be written
\bea \label{coordtrans}
x^{\mu} \mapsto \tilde{x}^{\mu} &=& e^{\xi^{\alpha} \partial_{\alpha}} x^{\mu} 
\, ,
\eea 
where $\xi^{\mu}$ is known as the ``gauge generator'', and is a small quantity in the perturbative expansion. A transformation of this type leaves all background quantities invariant, but changes the form of the perturbations. In this expression we have used the exponential map between coordinates systems, which guarantees that the group structure of the manifold is preserved. The explicit form of the transformation that should be applied to a tensor, $\mathcal{T}$, under the map presented in Eq. (\ref{coordtrans}), is given by
\bea \label{expmap} 
\tilde{\mathcal{T}} =  e^{\mathcal{L}_{\xi}} \mathcal{T} 
= \mathcal{T} + \mathcal{L}_{\xi}\mathcal{T} + {\textstyle\frac{1}{2}} \mathcal{L}_{\xi}^2 \mathcal{T} + \ldots
\, ,
\eea
where $\tilde{\mathcal{T}}$ is the transformed tensor and $\mathcal{L}_{\xi}$ is the Lie derivative along $\xi^{\mu}$. For a rank-2 tensor, $\mathcal{T}$, the Lie derivative is given by
\bea
\mathcal{L}_{\xi}\mathcal{T}_{\mu \nu} \equiv \mathcal{T}_{\mu \lambda}\xi^{\lambda}_{,\nu} + \mathcal{T}_{\lambda \nu}\xi^{\lambda}_{,\mu} + \mathcal{T}_{\mu \nu , \lambda}\xi^{\lambda}.
\eea
With Eqs. (\ref{coordtrans}), (\ref{expmap}) and the perturbed tensor $\mathcal{T}$ in hand, we can specify how the gauge generator $\xi^{\mu}$ should be expanded in orders of smallness, and then calculate the corresponding transformation of $\mathcal{T}$ order-by-order in the perturbations.

In principle, when expanding the gauge generator $\xi^{\mu}$ one could include terms at any order possible in the parameters $\epsilon$ and $\eta$. This, however, is not strictly necessary, as some orders will serve to produce new terms in the tensor $\tilde{\mathcal{T}}$ that are of no physical interest. This is the same type of problem that occurred when we expanded the metric in Section \ref{sec:pertexp}. The terms we wish to retain in $\xi^{\mu}$, and their orders of magnitude, are given by the following expressions:
\bea 
\xi^{0} &=& \xi^{(1, 0)0} + \xi^{(0, 3)0} + \xi^{(1, 2)0} + \ldots \sim \epsilon L_C + \eta^3 L_N + \epsilon \eta^3 L_N + \ldots  \label{ourxi0} \\[5pt] \xi^{i} &=& \xi^{(1, 0)i} + \xi^{(0, 2)i} + \xi^{(1,1 )i}+ \xi^{(1, 2)i} +  {\textstyle\frac{1}{2}} \xi^{(0, 4)i} +\ldots  \sim \epsilon L_C + \eta^2 L_N + \epsilon \eta^2 L_N + \eta^4 L_N + \ldots \,  . \label{ourxii}
\eea
We make several comments on these expressions. Firstly, one may note that each of the terms is proportional to a length scale, this is because the gauge generator $\xi^{\mu}$ corresponds to a change in space-time coordinates $x^{\mu}$ and coordinates have dimensions of length. The particular length scale assigned to each term is done in the same way as described in Section \ref{sec:pertexp}. Secondly, one may also note that while terms of $\mathcal{O}(\epsilon L_C)$ appear similarly in both $\xi^0$ and $\xi^i$, the order of terms perturbed in the parameter $\eta$ appear at different orders in $\xi^0$ and $\xi^i$. This is, once again, because time and space derivatives on cosmologically perturbed quantities add the same order of smallness whereas they add different orders of smallness in post-Newtonian perturbation theory. The ellipses in Eqs. (\ref{ourxi0}) and (\ref{ourxii}) correspond to terms that are smaller than those required to transform the field equations presented in Section \ref{sec:fe}. 

The lowest-order-cosmological gauge generators, $\xi^{(1,0)\mu}$, are of exactly the same order as the ones used in normal cosmological perturbation theory at linear order. These are the parts of the gauge generator that will create metric perturbations at order $g_{\mu \nu}^{(1,0)}$, in the usual way. This is just what we expect, as our cosmological metric perturbations are, for all intents and purposes, exactly the same as those used in standard cosmological perturbation theory ({\it i.e.} they have the same size, and vary in the same way in space and time). Additionally, the post-Newtonian gauge generators $\xi^{(0,3)0}$, $\xi^{(0,2)i}$ and $\xi^{(0,4)i}$ are exactly the same as those that occur in usual post-Newtonian perturbation theory \cite{poisson}. All mixed order gauge generators are unique to our two parameter expansion, and have no counterpart in either standard cosmological perturbation theory or standard post-Newtonian theory.

We formed the above gauge generators, Eq. (\ref{ourxi0}) and (\ref{ourxii}), in the same way as the perturbed energy density and metric, refer to Section \ref{pert2para}, such that the gauge generator contains the minimum number of perturbations necessary for a two-parameter system. We wrote an initial ansatz gauge generator with care because of the different length scales involved. The initial ansatz was given by the sum of the gauge generators used in cosmological perturbation theory, post-Newtonian gravity and mixed order gauge generators, that are products of the lowest-order gauge generators in both the cosmological and the post-Newtonian sectors, this gives $\xi^{(1,3)0}$ and $\xi^{(1,2)i}$. However, the terms in the final ansatz metric, given in Section \ref{pert2para}, strictly imply we require gauge generators of order $\xi^{(1,1)i}$ and $\xi^{(1,2)0}$ because we want to find and transform along \textit{all} possible degrees of freedom\footnote{The $ij$ and $0i$ parts of our initial metric ansatz produced new potentials of $\mathcal{O}(\epsilon \eta)$ and $\mathcal{O}(\epsilon \eta^2)$, respectively. As explained in Section \ref{pert2para}, we therefore included the extra metric components $g^{(1,1)}_{ij}$ and $g^{(1,2)}_{0i}$ in our new ansatz metric. The existence of these potentials then implies that we should have gauge generators of order $\xi^{(1,1)i}$ and $\xi^{(1,2)0}$, as we we want to find and transform along \textit{all} possible degrees of freedom.}. Therefore, we also include gauge generators $\xi^{(1,1)i}$ and $\xi^{(1,2)0}$ in our new ansatz gauge generator, given by Eqs. (\ref{ourxi0}) and (\ref{ourxii}). Now this gauge generator has the minimal number of perturbations necessary to create all necessary transformations to the metric, and stress-energy tensor.

By substitution of Eqs. (\ref{ourxi0}) and (\ref{ourxii}), into Eqs. (\ref{coordtrans}) and (\ref{expmap}), we can calculate how the metric and energy-momentum tensors transform under these infinitesimal coordinate transformations, order-by-order in perturbations. The rest of this section presents these results in detail.
 
\subsection{Transformation of the metric}

We begin by transforming the different components of the metric using
\bea
\tilde{g}_{\mu \nu} = g_{\mu \nu} + \mathcal{L}_{\xi}g_{\mu \nu} + {\textstyle\frac{1}{2}} \mathcal{L}_{\xi}^2 g_{\mu \nu} + \ldots \label{expmapg} \, , 
\eea
which is given by the exponential map, Eq. (\ref{expmap}), and where the expansion of the gauge generator $\xi^{\mu}$ is given by Eqs. (\ref{ourxi0}) and (\ref{ourxii}). Having done this, we will then proceed to perform an invariant decomposition of the results, so we split the metric into scalar, divergenceless vector, and transverse and trace-free tensor parts. This will be useful for constructing gauge invariant quantities, and writing down the governing equations, in Section \ref{sec:gaugeinvariants}. Throughout this section we will assume $L_N/L_C \sim \eta$, as in Section \ref{sec:fe}, but not $\epsilon \sim \eta^2$. 

\subsubsection{Transformation of metric components}

\noindent
{\bf The time-time component:} The perturbations of the time-time component of the metric, up to the order we wish to consider here, transform under the exponential map in Eq. (\ref{expmapg}) in the following way:
\bea  
h_{00}^{(0,2)} &\mapsto & \tilde{h}_{00}^{(0,2)} = h_{00}^{(0,2)} \label{h0002trans} \\[5pt] 
h_{00}^{(1,0)} &\mapsto & \tilde{h}_{00}^{(1,0)} = h_{00}^{(1,0)} - 2 \dot{\xi}^{(1, 0)0} \label{h0010trans} \\[5pt] 
h_{00}^{(1,1)} &\mapsto & \tilde{h}_{00}^{(1,1)} = h_{00}^{(1,1)} + h_{00,i}^{(0,2)} \xi^{(1,0)i} \label{h0011trans} \\[5pt]
h_{00}^{(1,2)} &\mapsto & \tilde{h}_{00}^{(1,2)} = h_{00}^{(1,2)} + \dot{h}_{00}^{(0,2)} \xi^{(1,0)0}    +2h_{00}^{(0,2)} \dot{\xi}^{(1,0)0} \label{h0012trans} \\[5pt] 
h_{00}^{(0,4)} & \mapsto & \tilde{h}_{00}^{(0,4)} = h_{00}^{(0,4)} - 4 \dot{\xi}^{(0,3)0} + 2h_{00,i}^{(0,2)}\xi^{(0,2)i} \, .\label{h0004trans}
\eea

We note that in addition to these transformations, each of which contains terms with the same order-of-magnitude, there is also a term generated from Eq. (\ref{expmapg}) in this component of the metric that is 
\bea
\label{anom1}
{\textstyle\frac{1}{2}}h_{00,ij}^{(0,2)} \xi^{(1,0)i} \xi^{(1,0)j} \, ,
\eea
which is of the $\mathcal{O}(\epsilon^2)$ when the length scales are taken into account appropriately. However, this term appears in the $\mathcal{O}(\eta^4L_N^{-2})$ $00$-field equation, Eq. (\ref{e00PN}), in the form of $R^{(2,0)}_{\mu \nu} \sim {\textstyle\frac{1}{2}} \nabla^2(h_{00,ij}^{(0,2)} \xi^{(1,0)i} \xi^{(1,0)j}) \sim {\textstyle\frac{1}{2}} \nabla^2h_{00,ij}^{(0,2)} \xi^{(1,0)i} \xi^{(1,0)j} \sim \epsilon^2L_N^{-2} \sim \eta^4L_N^{-2}$, when $\epsilon \sim \eta^2$. We discuss how such a term cancels with another term in the field equations in Section \ref{sec:FinalFieldEqs}.

\vspace{10pt}
\noindent
{\bf The time-space components:} The perturbations of the time-space parts of the metric transform, according to Eq. (\ref{expmapg}), in the following way:
\bea 
h_{0i}^{(0,3)}  \mapsto  \tilde{h}_{0i}^{(0,3)} &=& h_{0i}^{(0,3)} - \frac{1}{a} \xi^{(0, 3)0}_{,i} + a \dot{\xi}^{(0,2)}_{i}  \label{h0i03trans} \\[5pt] 
h_{0i}^{(1,0)} \mapsto  \tilde{h}_{0i}^{(1,0)} &=& h_{0i}^{(1,0)} - \frac{1}{a} \xi^{(1, 0)0}_{,i}+ a \dot{\xi}^{(1, 0)}_{i}  \label{h0i10trans} \\
h_{0i}^{(1,2)}  \mapsto   \tilde{h}_{0i}^{(1,2)} &=& h_{0i}^{(1,2)}  -\frac{1}{a}\dot{\xi}^{(1,2)0}_{,i}+ a \dot{\xi}^{(1,1)}_{i} +\chi^{(1,2)}_i \, ,
\eea
where we define
\bea
\chi^{(1,2)}_i &\equiv&\frac{1}{a}h_{00}^{(0,2)}\xi^{(1,0)0}_{,i} + a\left( h^{(0,2)}_{ij} +  \xi^{(0,2)}_{(i,j)} \right)  \dot{\xi}^{(1,0)j} + \left( h_{0i}^{(0,3)} - \frac{1}{2a} \xi^{(0,3)0}_{,i}  +\frac{1}{2}a \dot{\xi}^{(0,2)}_i \right)_{,j}\xi^{(1,0)j} \nonumber \\ 
&&+ \left(h_{0j}^{(1,0)} -\frac{1}{2a}\xi^{(1,0)0}_{,j} +\frac{1}{2}a \dot{\xi}^{(1,0)}_j \right)\xi^{(0,2)j}_{,i}            \, . \label{h0i12trans} 
\eea

\vspace{10pt}
\noindent
{\bf The space-space components:} The transformations of the perturbations in the space-space part of the metric are more lengthy than the previous cases. They transform under the exponential map in Eq. (\ref{expmapg}) in the following way:
\bea  
h_{ij}^{(0,2)} & \mapsto & \tilde{h}_{ij}^{(0,2)}= h_{ij}^{(0,2)} +  2\xi^{(0,2)}_{(i,j)}    \label{hij02trans}\\[5pt] 
h_{ij}^{(1,0)} & \mapsto & \tilde{h}_{ij}^{(1,0)} = h_{ij}^{(1,0)} + 2 \frac{\dot{a}}{a}  \xi^{(1, 0)0}\delta_{ij}  + 2\xi^{(1, 0)}_{(i,j)}   \label{hij10trans}\\[5pt]
h_{ij}^{(1,1)} & \mapsto & \tilde{h}_{ij}^{(1,1)} =  h_{ij}^{(1,1)} + 2\xi^{(1,1)}_{(i,j)} +\chi^{(1,1)}_{ij}  \label{hij11trans}\\[5pt]
h_{ij}^{(1,2)} & \mapsto & \tilde{h}_{ij}^{(1,2)} =  h_{ij}^{(1,2)} + 2\xi^{(1,2)}_{(i,j)} +\chi^{(1,2)}_{ij}  \label{hij12trans}\\[5pt]
h_{ij}^{(0,4)} & \mapsto & \tilde{h}_{ij}^{(0,4)} = h_{ij}^{(0,4)} + 4 \frac{\dot{a}}{a} \xi^{(0,3)0}\delta_{ij} + 2\xi^{(0,4)}_{(i,j)}  + \chi_{ij}^{(0,4)} \,  , \label{hij04trans}
\eea
where $\chi_{ij}^{(1,1)}$, $\chi_{ij}^{(1,2)}$ and $\chi_{ij}^{(0,4)}$ are defined as 
\bea
\chi^{(1,1)}_{ij} &\equiv & \left(h^{(0,2)}_{ij} + 2\xi^{(0,2)}_{(i,j)} \right)_{,k}\xi^{(1,0)k}  \\
\chi^{(1,2)}_{ij} &\equiv & \left(h_{ij}^{(0,2)} + \xi^{(0,2)}_{(i,j)} \right)\dot{•} \; \xi^{(1,0)0} + 2 \frac{\dot{a}}{a}\left(h_{ij}^{(0,2)} +2 \xi^{(0,2)}_{(i,j)} \right)\xi^{(1,0)0} 
+ \left(h^{(0,2)}_{ik} +\xi^{(0,2)}_{(i,k)} \right)\xi^{(1,0)k}_{,j}  \nonumber \\[5pt]
&& + \left(h^{(0,2)}_{jk} +\xi^{(0,2)}_{(j,k)} \right)\xi^{(1,0)k}_{,i} + \left(h^{(1,0)}_{ik} +\xi^{(1,0)}_{(i,k)} \right)\xi^{(0,2)k}_{,j} + \left(h^{(1,0)}_{jk} +\xi^{(1,0)}_{(j,k)} \right)\xi^{(0,2)k}_{,i}  \\[5pt] 
\chi_{ij}^{(0,4)} &\equiv & 2\left(h^{(0,2)}_{ij} + \xi^{(0,2)}_{(i,j)} \right)_{,k}\xi^{(0,2)k} + 2\left(h^{(0,2)}_{ik} +\xi^{(0,2)}_{(i,k)} \right)\xi^{(0,2)k}_{,j} + 2\left(h^{(0,2)}_{jk} +\xi^{(0,2)}_{(j,k)} \right)\xi^{(0,2)k}_{,i} \label{chi04ij} \, .
\eea

Before finishing this section, let us comment on the dependence of some of these terms on the condition $L_N/L_C \sim \eta$. In the time-time transformation the only terms that depend on this relation are $h^{(0,2)}_{00,i}\xi^{(1,0)i}$ and $\dot{h}_{00}^{(0,2)}\xi^{(1,0)0}$, which, once length scales are taken into account properly, appear at $\mathcal{O}(\epsilon \eta)$ and $\mathcal{O}(\epsilon \eta^2)$, respectively. If a different relationship between $L_N$ and $L_C$ had been chosen then this term would have appeared at a different order, and could appear in any equation greater than or equal to $\epsilon \eta$ and $\epsilon \eta^2$, respectively, before violating the bound in Eq. (\ref{xleqeta}). Similarly, in the transformation of the time-space and space-space components of the metric some of the terms in $\chi^{(1,2)}_i$ and $\chi^{(1,2)}_{ij}$, and terms $4\frac{\dot{a}}{a}\xi^{(0,3)0}\delta_{ij}$ and $\chi^{(1,1)}_{ij}$, all depend on the relationship between $L_N$ and $L_C$, and would appear at different orders if a different choice had been made for these length scales. 

\subsubsection{Transformation of irreducibly-decomposed potentials}

Having performed the gauge transformation of our metric components, in the previous section, we can now perform an irreducible decomposition of these objects into scalars, divergenceless vectors ($V^i_{\phantom{i} , i}=0$), and transverse and trace-free tensors ($h^i_{\phantom{i} i}=0$ and $h^{ij}_{\phantom{ij},j}=0$). These are the quantities that are most often considered in cosmological perturbation theory, and that usually decouple from each at first order in perturbations. We decompose our metric potentials into these variables in the following way, omitting superscripts for simplicity:
\be
h_{00}  \equiv  \phi \, , \qquad h_{0i} \equiv B_{,i} + B_i \qquad {\rm and} \qquad h_{ij} \equiv  - \psi\delta_{ij} + E_{,ij} + F_{(i,j)} + {\textstyle\frac{1}{2}}\hat{h}_{ij} \, .
\ee
Similarly, our gauge generators will be decomposed such that 
\be
\xi^0 \equiv \delta t \qquad {\rm and} \qquad \xi^i \equiv 
\delta x^{,i} + \delta x^i \, .
\ee
We will now present the result of gauge transformations on each of the irreducibly decomposed objects, in each of the sectors of our perturbation theory.

\vspace{10pt}
\noindent
{\bf Cosmological scalar, vector and tensor potentials:}
The gauge transformations given in Eqs. (\ref{h0010trans}), (\ref{h0i10trans}), and (\ref{hij10trans}) now allow us to write down the transformation of the decomposed metric components in the cosmological sector of our theory. For the scalar potentials these transformations are given by
\bea
\tilde{\phi}^{(1,0)} & = & \phi^{(1,0)} - 2\dot{\delta t}^{(1, 0)} \sim \epsilon  \label{phi10trans} \\[5pt]
\tilde{\psi}^{(1,0)} & = & \psi^{(1,0)} - 2 \frac{\dot{a}}{a} \delta t^{(1,0)} \sim \epsilon  \\[5pt]
\tilde{B}^{(1,0)} & = & B^{(1,0)} + a \dot{\delta x}^{(1,0)} - \frac{1}{a} \delta t^{(1, 0)} \sim \epsilon \eta^{-1} L_N  \\[5pt]
\tilde{E}^{(1, 0)} & = & E^{(1, 0)} + 2 \delta x^{(1, 0)} \sim \epsilon \eta^{-2} L_N^2 \, ,
\eea
for the vector potentials they are
\bea
\tilde{B}_i^{(1, 0)} & = & B_i^{(1, 0)} + a \dot{\delta x_i}^{(1, 0)} \sim \epsilon  \\[5pt]
\tilde{F}_i^{(1, 0)} & = & F_i^{(1, 0)} + 2 \delta x_i^{(1,0)} \sim \epsilon \eta^{-1} L_N \, ,
\eea
and for the tensor potential this transformation is
\bea
\tilde{\hat{h}}_{ij}^{(1,0)} & = & \hat{h}_{ij}^{(1,0)} \sim \epsilon \, . \label{hijtens10trans}
\eea
As in previous equations, the quantity after the $\sim$ sign gives the order of each of these potentials in terms of $\epsilon$, $\eta$ and any relevant length scales. We observe that the transformation of the above cosmological scalar, vector and tensor potentials in our two-parameter formalism are the same as those derived from linear cosmological perturbation theory \cite{malik}, perturbed in one parameter.

\vspace{10pt}
\noindent
{\bf Post-Newtonian scalar, vector and tensor potentials:}
The results given in Eqs. (\ref{h0002trans}), (\ref{h0004trans}), (\ref{h0i03trans}), (\ref{hij02trans}), and (\ref{hij04trans}) allow us to write the transformation of the decomposed post-Newtonian potentials. The scalar parts of the post-Newtonian potentials transform as
\bea
\tilde{\phi}^{(0,2)} &=& \phi^{(0,2)} \sim \eta^2  \\[5pt]
\tilde{\phi} ^{(0,4)}& =& \phi^{(0,4)} - 4 \dot{\delta t}^{(0,3)}  + 2 \phi^{(0,2)}_{,i} \left(\delta x^{(0,2),i} + \delta x^{(0,2)i}\right)   \sim  \eta^4 \label{eq00ScalarPhi04} \\[5pt]
\tilde{\psi}^{(0,2)} &=& \psi^{(0,2)} \sim \eta^2  \\[5pt]
\tilde{\psi}^{(0,4)} &=& \psi^{(0,4)} -4 \frac{\dot{a}}{a}\delta t^{(0,3)} + \frac{1}{2}\left(\nabla^{-2}\chi_{ij}^{(0,4),ij} -\chi^{(0,4)}\right) \sim \eta^4  \\[5pt]
\tilde{B}^{(0,3)} &=& B^{(0,3)} + a \dot{\delta x}^{(0,2)} - \frac{1}{a} \delta t^{(0,3)} \sim \eta^3 L_N \\[5pt]
\tilde{E}^{(0,2)} &=& E^{(0,2)} + 2 \delta x^{(0,2)} \sim \eta^2 L_N^2  \\[5pt] 
\tilde{E}^{(0,4)} &=& E^{(0,4)} + 2 \delta x^{(0,4)}   
 + \frac{1}{2}\nabla^{-2}\left(3 \nabla^{-2} \chi^{(0,4),ij}_{ij} - \chi^{(0,4)}\right)   \sim  \eta^4 L_N^2 \, , 
\eea
the vector potentials transform as
\bea
\tilde{B}_i^{(0,3)} &=& B_i^{(0,3)} + a \dot{\delta x_i}^{(0,2)} \sim \eta^3 \\[5pt]
\tilde{F}_i^{(0,2)} &=& F_i^{(0,2)} + 2 \delta x_i^{(0,2)}\sim \eta^2 L_N   \\[5pt]
\tilde{F}_i^{(0,4)} &=& F_i^{(0,4)} + 2 \delta x_i^{(0,4)} + 2\nabla^{-2}\left(\chi^{(0,4),k}_{ik} - \nabla^{-2}\chi^{(0,4),kj}_{kj,i} \right)\sim  \eta^4 L_N \, ,
\eea
and the tensor potentials transform as
\bea
\tilde{\hat{h}}_{ij}^{(0,2)} &=& \hat{h}_{ij}^{(0,2)}\sim \eta^2 \\[5pt]
\tilde{\hat{h}}_{ij}^{(0,4)} &=& \hat{h}_{ij}^{(0,4)} + 2\chi^{(0,4)}_{ij} -4\nabla^{-2}\chi^{(0,4),k}_{k(i,j)} +\left(\nabla^{-2}\chi_{kl}^{(0,4),kl} - \chi^{(0,4)}\right)\delta_{ij} + \nabla^{-2}\left(\nabla^{-2}\chi^{(0,4),kl}_{kl} + \chi^{(0,4)}\right)_{,ij} \sim  \eta^4 \, . 
\eea
The quantity $\chi^{(0,4)}_{ij}$ is defined in Eq. (\ref{chi04ij}), and here we have written $\chi^{(n,m)} \equiv \delta ^{ij}\chi^{(n,m)}_{ij}$. In terms of irreducibly decomposed potentials, this quantity can be written as
\bea
\chi^{(0,4)}_{ij} &=& 2\left(-\psi^{(0,2)}_{,k}\delta_{ij} + E^{(0,2)}_{,ijk} + F^{(0,2)}_{(i,j)k} + \frac{1}{2}\hat{h}^{(0,2)}_{ij,k} + \delta x^{(0,2)}_{,ijk} + \delta x^{(0,2)}_{(i,j)k} \right)\left(\delta x^{(0,2),k} +\delta x^{(0,2)k} \right) \nonumber \\[5pt]
&&+ 2\left(-\psi^{(0,2)}\delta_{ik} + E^{(0,2)}_{,ik} + F^{(0,2)}_{(i,k)} + \frac{1}{2}\hat{h}^{(0,2)}_{ik} +\delta x^{(0,2)}_{,ik} + \delta x^{(0,2)}_{(i,k)} \right)\left(\delta x^{(0,2),k}_{,j} + \delta x^{(0,2)k}_{,j}  \right) \nonumber \\[5pt]
&&+ 2\left(-\psi^{(0,2)}\delta_{jk} + E^{(0,2)}_{,jk} + F^{(0,2)}_{(j,k)} + \frac{1}{2}\hat{h}^{(0,2)}_{jk} + \delta x^{(0,2)}_{,jk} + \delta x^{(0,2)}_{(j,k)} \right)\left(\delta x^{(0,2),k}_{,i} + \delta x^{(0,2)k}_{,i}  \right) \, .
\eea
This completes the full set of transformations in the post-Newtonian sector. We note that the lowest order post-Newtonian metric potentials $\phi^{(0,2)}$ and $\psi^{(0,2)}$ transform as expected from post-Newtonian gravity \cite{will}. As far as we are aware, the transformation of scalar, vector and tensor post-Newtonian potentials has not been calculated before. The above transformations are derived from our two-parameter formalism, but because there are only post-Newtonian (not cosmological or mixed-order) potentials and gauge generators in these transformations they also hold for one-parameter post-Newtonian gravity.

\vspace{10pt}
\noindent
{\bf Mixed-order scalar, vector and tensor potentials:}
The scalar part of the mixed-order potentials, up to the order considered in the field equations in Section \ref{sec:fe}, $\mathcal{O}(\epsilon \eta^2)$, transform in the following way:
\bea
\tilde{\phi}^{(1,1)} &=& \phi^{(1,1)} + \phi^{(0,2)}_{,i} \left(\delta x^{(1,0),i} + \delta x^{(1,0)i}\right)\sim \epsilon \eta  \\[5pt]
\tilde{\phi}^{(1,2)} &=& \phi^{(1,2)} + \dot{\phi}^{(0,2)} \delta t^{(1,0)} + 2 \phi^{(0,2)}\dot{\delta t}^{(1,0)} \sim \epsilon \eta^2 \\[5pt]
\tilde{\psi}^{(1,1)} &=& \psi^{(1,1)}  + \frac{1}{2}\left(\nabla^{-2}\chi_{ij}^{(1,1),ij} -\chi^{(1,1)}\right) \sim \epsilon \eta   \\[5pt]
\tilde{\psi}^{(1,2)} &=& \psi^{(1,2)} + \nabla^{-2}\left(\chi^{(1,2),k]l}_{k[l} + 2 \mathcal{C}_{k[l\vert ,m}^{,\vert k]}\mathcal{I}^{m,l}\right) \sim \epsilon \eta^2  \\[5pt]
\tilde{B}^{(1,2)} &=& B^{(1,2)} + a \dot{\delta x}^{(1,1)} -\frac{1}{a}\delta t^{(1,2)}+ \nabla^{-2}\chi^{(1,2),i}_i  \sim \epsilon \eta^2 L_N   \\[5pt]
\tilde{E}^{(1,1)} &=& E^{(1,1)}+ 2 \delta x^{(1,1)}
 + \frac{1}{2}\nabla^{-2}(3 \nabla^{-2} \chi^{(1,1),ij}_{ij} - \chi^{(1,1)})   \sim  \epsilon \eta L_N^2  \\
\tilde{E}^{(1,2)} &=& E^{(1,2)}+ 2 \delta x^{(1,2)} +\frac{1}{2}\nabla ^{-2} \left( \nabla ^{-2}\left(3\chi^{(1,2),kl}_{kl} + 6 \mathcal{C}_{kl,m}^{,k}\mathcal{I}^{m,l} - 2 \mathcal{C}^k_{k,l}\mathcal{I}^{m,l}\right) - \chi^{(1,2)} \right) \sim \epsilon \eta^2 L_N^2\, ,
\eea
where we have used anti-symmetric square brackets that are defined by $2\mathcal{T}_{[ij]} \equiv \mathcal{T}_{ij} - \mathcal{T}_{ji}$. The vector parts transform as
\bea
\tilde{B}^{(1,2)}_i&=& B^{(1,2)}_i + a \dot{\delta x}^{(1,1)}_i +\chi^{(1,2)}_i -\nabla^{-2}\chi^{(1,2),j}_{j,i}   \sim \epsilon \eta^2    \\
\tilde{F}^{(1,1)}_i &=& F^{(1,1)}_i  + 2 \delta x_i^{(1,1)} + 2\nabla^{-2}\left(\chi^{(1,1),k}_{ik} - \nabla^{-2}\chi^{(1,1),kj}_{kj,i} \right)\sim  \epsilon \eta L_N  \\
\tilde{F}^{(1,2)}_{i}&=& F^{(1,2)}_{i} + 2 \delta x^{(1,2)}_{i}   -2 \nabla^{-2}\nabla^{-2}\left( 2\chi^{(1,2),kl}_{k[i,l]} -4 \mathcal{C}^{,k}_{k[i,l]m}\mathcal{I}^{m,l} - \nabla^2\mathcal{C}_{ki,m}\mathcal{I}^{m,k} + \mathcal{C}^{,kl}_{kl,m} \mathcal{I}^m_{,i} \right) \sim \epsilon \eta^2 L_N\, ,
\eea
and the tensor parts transform as
\bea
\tilde{\hat{h}}^{(1,1)}_{ij} &=& \hat{h}^{(1,1)}_{ij} + 2 \chi^{(1,1)}_{ij}  - 4\nabla^{-2}\chi^{(1,1),k}_{k(i,j)} + \nabla^{-2} \chi^{(1,1),kl}_{kl} \delta_{ij} - \chi^{(1,1)} \delta_{ij} 
   + \nabla^{-2}\nabla^{-2}\chi^{(1,1),kl}_{kl,ij} + \nabla^{-2}\chi^{(1,1)}_{,ij} \sim \epsilon \eta  \\
\tilde{\hat{h}}^{(1,2)}_{ij} &=& \hat{h}^{(1,2)}_{ij} + 2 \chi^{(1,2)}_{ij}  - 4\nabla^{-2}\chi^{(1,2),k}_{k(i,j)} + \nabla^{-2} \chi^{(1,2),kl}_{kl} \delta_{ij} - \chi^{(1,2)} \delta_{ij} 
   + \nabla^{-2}\nabla^{-2}\chi^{(1,2),kl}_{kl,ij} + \nabla^{-2}\chi^{(1,2)}_{,ij} \nonumber \\ 
   && + 4 \nabla^{-2} \nabla^{-2} \left(  \nabla^2 \mathcal{C}_{ij,mk} \mathcal{I}^{m,k} - \nabla^2 \C_{k(i,j)m}\I^{m,k} -2\C_{k(i,j)klm}\I^{m,l} -\nabla^2 \C_{k(i \vert, m}^{,k} \I^m_{,\vert j)} + \C^{,k( l \vert }_{kl,mn}\I^{m, \vert n)}\delta_{ij}   \right) \nonumber \\
   && +\nabla^{-2} \nabla^{-2}\left( - \nabla^2 \C^k_{k,ml} \I^{m,l} \delta_{ij} +2\C_{kl,mij}^{,k}\I^{m,l} + 2 \C^{,kl}_{kl,m(i} \I^m_{,j)} + 2\C_{ij,mk} \I^{m,k}      \right) \sim \epsilon \eta^2 \, .
\eea
Note that in the above equations we define $\nabla^{-2}f(\chi^{(n,m)})$ such that $\nabla^2[\nabla^{-2}f(\chi^{(n,m)})]$ is the leading order part of $f(\chi^{(n,m)})$ and no smaller, which strictly excludes higher order terms in $f(\chi^{(n,m)})$. In the above equations we have written $\chi^{(1,2)}_{i}$, $\chi^{(1,2)}_{ij}$ and $\chi^{(1,1)}_{ij}$ in terms of scalar, vector and tensor potentials and $\chi^{(1,1)}_{ij}$ in terms of $\C_{ij,m} $ and $ \I^{m}$ in the following way
\bea
\chi^{(1,2)}_{i} &=& \frac{1}{a}\phi^{(0,2)}\delta t^{(1,0)}_{,i} + a \left(-\psi^{0,2}\delta_{ij} +E^{(0,2)}_{,ij} +F^{(0,2)}_{(i,j)} + \frac{1}{2}\hat{h}^{(0,2)}_{ij} + \delta x^{(0,2)}_{,ij} + \delta
x^{(0,2)}_{(i,j)}  \right)\left(\delta x^{(1,0),j} +\delta x^{(1,0)j} \right)\dot{•}   \\
&& + \left(B^{(0,3)}_{,i}+B^{(0,3)}_{i} -\frac{1}{2a} \delta t^{(0,3)}_{,i} +\frac{a}{2}\left(\delta x^{(0,2)}_{,i} +\delta x^{(0,2)}_{i} \right)\dot{•}  \right)_{,j} \left(\delta x^{(1,0),j} + \delta x^{(1,0)j}\right) \nonumber \\
&& + \left(B^{(1,0)}_{,j}+B^{(1,0)}_{j} -\frac{1}{2a} \delta t^{(1,0)}_{,j} +\frac{a}{2}\left(\delta x^{(1,0)}_{,j} +\delta x^{(1,0)}_{j} \right)\dot{•}  \right) \left(\delta x^{(0,2),j} + \delta x^{(0,2)j}\right)_{,i} \nonumber 
\eea
\bea
\chi^{(1,2)}_{ij} &=& \left(-\psi^{(0,2)} \delta_{ij} + E^{(0,2)}_{,ij} + F^{(0,2)}_{(i,j)} + \frac{1}{2}\hat{h}^{(0,2)}_{ij} + \delta x^{(0,2)}_{,ij} + \delta x^{(0,2)}_{(i,j)} \right)\dot{•} \, \delta t^{(1,0)} \nonumber \\[5pt]
&\quad & + 2\frac{\dot{a}}{a} \left( -\psi^{(0,2)} \delta_{ij} + E^{(0,2)}_{,ij} + F^{(0,2)}_{(i,j)} + \frac{1}{2}\hat{h}^{(0,2)}_{ij} + 2\delta x^{(0,2)}_{,ij} + 2\delta x^{(0,2)}_{(i,j)} \right)\delta t^{(1,0)} \nonumber \\[5pt]
&\quad & + \left(-\psi^{(0,2)} \delta_{ik} + E^{(0,2)}_{,ik} + F^{(0,2)}_{(i,k)} + \frac{1}{2}\hat{h}^{(0,2)}_{ik} + \delta x^{(0,2)}_{,ik} + \delta x^{(0,2)}_{(i,k)} \right)\left(\delta x^{(1,0),k} + \delta x^{(1,0)k} \right)_{,j} \nonumber \\[5pt]
&\quad & + \left(-\psi^{(0,2)} \delta_{jk} + E^{(0,2)}_{,jk} + F^{(0,2)}_{(j,k)} + \frac{1}{2}\hat{h}^{(0,2)}_{jk} + \delta x^{(0,2)}_{,jk} + \delta x^{(0,2)}_{(j,k)} \right)\left(\delta x^{(1,0),k} + \delta x^{(1,0)k} \right)_{,i} \nonumber \\[5pt]
&\quad & + \left(-\psi^{(1,0)} \delta_{ik} + E^{(1,0)}_{,ik} + F^{(1,0)}_{(i,k)} + \frac{1}{2}\hat{h}^{(1,0)}_{ik} + \delta x^{(1,0)}_{,ik} + \delta x^{(1,0)}_{(i,k)} \right)\left(\delta x^{(0,2),k} + \delta x^{(0,2)k} \right)_{,j} \nonumber \\[5pt]
&\quad & + \left(-\psi^{(1,0)} \delta_{jk} + E^{(1,0)}_{,jk} + F^{(1,0)}_{(j,k)} + \frac{1}{2}\hat{h}^{(1,0)}_{jk} + \delta x^{(1,0)}_{,jk} + \delta x^{(1,0)}_{(j,k)} \right)\left(\delta x^{(0,2),k} + \delta x^{(0,2)k} \right)_{,i}  \\[5pt]
\chi^{(1,1)}_{ij} &=&  \C_{ij,k} \I^{k} \, , 
\eea
where we have defined
\bea
\C_{ij,k} & \equiv &  \left(-\psi^{(0,2)} \delta_{ij} + E^{(0,2)}_{,ij} + F^{(0,2)}_{(i,j)} + \frac{1}{2}\hat{h}^{(0,2)}_{ij} + \delta x^{(0,2)}_{,ij} + \delta x^{(0,2)}_{(i,j)} \right)_{,k} \sim \eta^2L_N^{-1}  \\
\I^k &\equiv & \delta x^{(1,0),k} +\delta x^{(1,0)k} \, \sim \epsilon \eta^{-1} L_N.
\eea
This completes our treatment of gauge transformations of the metric tensor. These transformations are original results and will be used in Section \ref{sec:gaugeinvariants} to construct gauge invariant potentials. Moreover, the transformation of our mixed-order quantities are purely a result of our two parameter formalism.

\subsection{Transformations of the energy-momentum tensor} \label{sec:transofTmunu}

The same freedoms, associated with infinitesimal coordinate transformations, can also be considered in the context of the energy-momentum tensor and its components. In the following we find how this tensor behaves under a gauge transformation of the form specified in Eq. (\ref{expmap}). As before, we will first calculate the explicit transformations that apply to the components of this tensor, and then to their irreducibly decomposed scalar, vector and tensor parts. Again, we take $L_N/L_C \sim \eta$ throughout this section.

\subsubsection{Transformation of the components of $T_{\mu \nu}$} \label{transTmunu}

\vspace{10pt}
\noindent
{\bf The transformation of ${\mathbf T_{00}}$:}
Using the exponential map in Eq. (\ref{expmap}), and the gauge generators specified in Eqs. (\ref{ourxi0}) and (\ref{ourxii}), we find the following transformations:
\bea
\tilde{\rho}^{(0,2)} &=& \rho^{(0,2)} \sim \frac{\eta^2}{L_N^2}  \label{T0200trans}\\[5pt] 
\tilde{\rho}^{(1,1)} &=& \rho^{(1,1)} + \rho^{(0,2)}_{,i}\xi^{(1,0)i} \sim \frac{\epsilon \eta}{L_N^2}  \label{T1100trans}\\[5pt]
\tilde{\rho}^{(1,0)} + \tilde{\rho}^{(1,2)} - \tilde{h}_{00}^{(1,0)}\tilde{\rho}^{(0,2)} &=& \rho^{(1,0)} + \rho^{(1,2)}  - h_{00}^{(1,0)}\rho^{(0,2)}  + \dot{\rho}^{(0,2)}\xi^{(1,0)0} + 2\rho^{(0,2)}\dot{\xi}^{(1,0)0}  \sim  \frac{\epsilon \eta^2}{L_N^2} \quad \\[5pt]
\frac{1}{2}\tilde{\rho}^{(0,4)} - \tilde{h}_{00}^{(0,2)} \tilde{\rho}^{(0,2)} +  \tilde{\rho}^{(0,2)}\tilde{v}^{(0,1)i} \tilde{v}_i^{(0,1)}  & = & \frac{1}{2} \rho^{(0,4)} - h_{00}^{(0,2)}\rho^{(0,2)} +  \rho^{(0,2)}v^{(0,1)i}v_i^{(0,1)}  + \rho^{(0,2)}_{,i} \xi^{(0,2)i}  \sim   \frac{\eta^4}{L_N^2} \, . \qquad \qquad 
\eea
We note that the Stewart-Walker lemma tells us $\rho^{(0,2)}$ is gauge invariant because there is no background energy density \cite{SWALKER}, which is exactly what we find.

Finally, we note that one further term is generated by the transformation of this part of the energy-momentum tensor: $T_{\mu \nu}^{(2,0)} \sim {\textstyle\frac{1}{2}} \rho^{(0,2)}_{,ij}\xi^{(1,0)i}\xi^{(1,0)j} \sim \epsilon^2L_N^{-2}$. This term would appear in the $\eta^4L_N^{-2}$ field equation along with $R^{(2,0)}_{\mu \nu} \sim \epsilon^2 L_N^{-2}$ (see Eq. (\ref{anom1})). We explain what happens to the terms of $\mathcal{O}(\epsilon^2L_N^{-2})$ in Section \ref{sec:FinalFieldEqs}. 

\vspace{10pt}
\noindent
{\bf The transformation of ${\mathbf T_{0i}}$:}
The same gauge transformations give the following results for the time-space components of the energy-momentum tensor:
\bea
-a \tilde{\rho}^{(0,2)}\tilde{v}_i^{(0,1)} &=& -a \rho^{(0,2)}v_i^{(0,1)} \sim \frac{\eta^3}{L_N^2} \\[5pt]
-a \tilde{\rho}^{(0,2)}\tilde{v}_i^{(1,0)} -a \tilde{\rho}^{(1,1)}\tilde{v}_i^{(0,1)} - a\tilde{\rho}^{(0,2)}\tilde{h}_{0i}^{(1,0)} &=& -a \rho^{(0,2)}v_i^{(1,0)}-a \rho^{(1,1)}v_i^{(0,1)} -a\rho^{(0,2)}h_{0i}^{(1,0)} \nonumber \\
&& + \rho^{(0,2)}\xi^{(1,0)0}_{,i} - a \left(\rho^{(0,2)}v^{(0,1)}_i\right)_{,j}\xi^{(1,0)j}  \sim  \frac{\epsilon \eta^2}{L_N^2} \, . \qquad 
\eea

\vspace{10pt}
\noindent
{\bf The transformation of ${\mathbf T_{ij}}$:}
Finally, the gauge transformation of the space-space components of the energy-momentum tensor gives
\bea
a^2\tilde{\rho}^{(0,2)}\tilde{v}_i^{(0,1)}\tilde{v}_j^{(0,1)} + a^2 \tilde{p}^{(0,4)} \delta_{ij} &=& a^2\rho^{(0,2)}v_i^{(0,1)}v_j^{(0,1)} + a^2 p^{(0,4)} \delta_{ij}  \sim \frac{\eta^4}{L_N^2} \\[5pt]
a^2 \tilde{p}^{(1,0)} \delta_{ij} &=& a^2 p^{(1,0)} \delta_{ij} \sim \frac{\epsilon \eta^2}{L_N^2} \, . \label{T10ijtrans}
\eea
Again, we note $p^{(1,0)}$ is gauge invariant because there is no homogeneous (or constant) background pressure. This is because at late times the Universe is dust dominated, but we allow for a small cosmological source of pressure.

\subsubsection{Transformation of scalar, vector and tensor parts of $T_{\mu\nu}$}

The irreducible decomposition of the quantities that appear in the energy-momentum tensor are simplified by the fact that they are all scalars, with the exception of the 3-velocity, $v_i$. This vector can be split into scalar and divergenceless vector parts as follows:
\be
v_i \equiv  v_{,i} + \hat{v}_i \, ,
\ee
where $\hat{v}^i_{\phantom{i} ,i}=0$. The scalar degrees of freedom in the metric are then given by $\rho$, $p$ and $v$, while the only divergenceless vector is given by $\hat{v}_i$. There are no transverse and trace-free tensorial terms in the stress-energy tensor, up to the order we consider, and as defined in Eq. (\ref{e:Tmunu}).

\vspace{10pt}
\noindent
{\bf Cosmological and mixed-order scalar and vector energy-momentum sources:}
Using Eq. (\ref{T0200trans})-(\ref{T10ijtrans}), we find that the irreducibly decomposed scalars and vectors in the cosmological sector transform according to
\bea
\tilde{\rho}^{(1,1)}&=&\rho^{(1,1)} +\rho^{(0,2)}_{,i}\left(\delta x^{(1,0),i} + \delta x^{(1,0)i}\right)  \\
\tilde{\rho}^{(1,0)} + \tilde{\rho}^{(1,2)} &=&  \rho^{(1,0)} + \rho^{(1,2)} + \dot{\rho}^{(0,2)}\delta t^{(1,0)}   \\[5pt]
\tilde{p}^{(1,0)} &=&  p^{(1,0)} \, ,
\eea
and
\be
\label{vcostrans}
\tilde{v}_i^{(1,0)} = v_i^{(1,0)} - a\dot{\xi}^{(1,0)}_i + v^{(0,1)}_{i,j} \xi^{(1,0)j} \, .
\ee
The scalar part of the 3-velocity, $v$, and the divergenceless vector part $\hat{v}_i$, can be found from taking the divergence of this last equation. We do not perform these operations explicitly here, as they result in less compact expressions. The quadratic term that appears in Eq. (\ref{vcostrans}) shows that the small-scale Newtonian velocity is important for determining how the large-scale velocity transforms. 

\vspace{10pt}
\noindent
{\bf Post-Newtonian scalar and vector energy-momentum sources:}
Eqs. (\ref{T0200trans})-(\ref{T10ijtrans}) can also be used to find the transformation of the scalar and vector parts of the post-Newtonian sector of our theory, which gives
\bea
\tilde{\rho}^{(0,2)} &=& \rho^{(0,2)}  \\[5pt]
\tilde{\rho}^{(0,4)} &=&  \rho^{(0,4)} + 2 \rho^{(0,2)}_{,i} \left(\delta x^{(0,2),i} + \delta x^{(0,2)i}\right)  \\[5pt]
\tilde{p}^{(0,4)} &=&  p^{(0,4)} \, , 
\eea
and
\be
\tilde{v}_i^{(0,1)} =  v_i^{(0,1)} \, .
\ee
This last equation states that both the scalar and vector parts of the 3-velocity are gauge invariant in this sector of the theory, at this order. The leading-order parts of the post-Newtonian energy density and pressure are also automatically gauge invariant. This is to be expected, as these equations describe Newtonian gravity at leading order, which of course transforms trivially under general coordinate transformations. These results differ from the quasi-static limit of cosmological perturbation theory, as space and time derivatives are treated differently and velocities come in at different orders \cite{Peebles}. This completes our study of the gauge transformations of this tensor.

\section{Constructing gauge invariant quantities}
\label{sec:gaugeinvariants}

Having performed infinitesimal coordinate transformations of the metric and energy-momentum-tensor, we are now in a position to isolate and remove the superfluous degrees of freedom associated with diffeomorphism covariance. This will leave us with a set of quantities that represent the physical degrees of freedom in the problem only, and will remove the possibility of any interference from spurious gauge modes.

Dealing with gauge freedoms can be done in a number of different ways, and is often approached differently in the respective literatures associated with post-Newtonian gravity \cite{will} and cosmological perturbation theory \cite{malik}. In post-Newtonian gravity, the usual method is to make a gauge choice by setting the sum of various parts of the perturbed field equations to zero. If suitable choices are made, and if they can be shown to be self-consistent, then this method can be used to remove all gauge freedom. This approach has the distinct benefit of allowing maximum simplification of the field equations, making these equations easier to solve, and the entire problem more tractable. However, it also has the drawback that one has to determine what is, or is not, a suitable choice of terms to eliminate from the field equations. This can sometimes be a challenge.

On the other hand, in the literature on cosmological perturbation theory a gauge choice is most usually made by irreducibly decomposing the metric and energy-momentum tensor, and then by setting some of the resulting terms to zero directly \cite{malik}. This leaves a more complicated set of field equations compared to post-Newtonian gravity, described in the previous paragraph, but does allow for the maximum possible simplification of the basic objects involved in the problem. Even in this case, however, it is still possible to leave behind residual gauge freedoms, if inappropriate choices are made. These problems were circumvented by Bardeen, who was the first to construct combinations of perturbations that remained invariant under general gauge transformations \cite{bardeen}. This removed all ambiguity, and allowed perturbed field equations to be written down that were guaranteed to be free from all gauge freedoms.

We choose to use the latter of these two approaches, to construct gauge invariant quantities associated with the perturbations to metric and energy-momentum tensors. This involves extending the method pioneered by Bardeen to post-Newtonian perturbations, as well as using some of the extensions of this method developed for use in second-order cosmological perturbation theory \cite{malik}. By the end of this section we will have written down gauge-invariant quantities for all of the perturbations described above, as well as the differential equations that govern them.

\subsection{Gauge-invariant metric perturbations}

Let us begin by constructing gauge-invariant quantities from the irreducibly decomposed metric tensor. The method we will use to do this is based on that developed for single-parameter cosmological perturbation theory \cite{malik}, and will be such that our gauge invariant quantities reduce to the metric perturbations in longitudinal gauge when $E=B=F_i = 0$ (we omit superscript indices here for simplicity). We note that other gauge choices are possible; we make this choice so that the field equations look similar to those in post-Newtonian gravity. The procedure we will use for this will be to choose gauge generators, $\delta x, \delta x^i $ and $\delta t$, such that $\tilde{E}= \tilde{B} =\tilde{F_i} =0$. We will then substitute these quantities back into the expressions for all of the transformed perturbations presented in Section \ref{Gchoice}. The results will be gauge invariant, as the original gauge transformations were written down in a completely arbitrary coordinate system. This means that newly constructed quantities cannot depend on any choice of gauge, and hence must be gauge invariant.

Below we present our results for the cosmological sector, the post-Newtonian sector, and the mixed-order sector of our expansion. All quantities have been checked, by explicit transformation, to ensure that they are in fact gauge invariant.

\vspace{10pt}
\noindent
{\bf Cosmological quantities:}
In the cosmological sector we find that we can form two independent scalar, one vector and one tensor gauge invariant quantities. These are given by:
\bea
\Phi^{(1,0)} &=& \phi^{(1,0)} -2a\dot{B}^{(1,0)} -2\dot{a}B^{(1,0)} + 2\dot{a}a\dot{E}^{(1,0)} + a^2\ddot{E}^{(1,0)}  \label{eqCosmoInvarPhi}  \\[5pt]
\Psi^{(1,0)} &=& \psi^{(1,0)} + \dot{a}a\dot{E}^{(1,0)} - 2\dot{a}B^{(1,0)}  \label{eqCosmoInvarPsi}  \\[5pt]
\mathbf{B}_i^{(1,0)} &=& B_i^{(1,0)} - \frac{a}{2}\dot{F}_i^{(1,0)}  \label{eqCosmoInvarPhiVect} \\[5pt]
\mathbf{h}_{ij}^{(1,0)} &=& \hat{h}_{ij}^{(1,0)} \, , \label{eqCosmoInvarhij}
\eea
which are all at $\mathcal{O}(\epsilon)$. These gauge invariant quantities are identical to those found by Bardeen, in the context of standard cosmological perturbation theory \cite{bardeen}. 

\vspace{10pt}
\noindent
{\bf Post-Newtonian quantities:}
In the post-Newtonian sector, at $\mathcal{O}(\eta^2)$, we can create two scalar, and one tensor, gauge invariant quantities:
\bea 
\Phi^{(0,2)} &=& \phi^{(0,2)}  \label{eqNewtPertInvar2} \\[5pt] 
\Psi^{(0,2)} &=& \psi^{(0,2)}   \label{eqPNinvarPsi} \\[5pt]
\mathbf{h}_{ij}^{(0,2)} &=& \hat{h}_{ij}^{(0,2)}  \label{eqPNinvarhij}  \, .
\eea
At $\mathcal{O}(\eta^3)$ there exists one gauge invariant vector,
\bea
\mathbf{B}_i^{(0,3)} &=& B_i^{(0,3)} - \frac{a}{2}\dot{F}_i^{(0,2)}  \label{eqPNinvarPhiVect} \, ,
\eea
while at $\mathcal{O}(\eta^4)$ there are two scalars and one tensor,
\bea
\Phi^{(0,4)} &=& \phi^{(0,4)} -4a\dot{B}^{(0,3)} -4\dot{a}B^{(0,3)} + 4\dot{a}a\dot{E}^{(0,2)} + 2a^2\ddot{E}^{(0,2)}  - \phi^{(0,2)},_i\left(E^{(0,2),i} + F^{(0,2)i}\right)  \label{eqPNinvarPhi} \\[5pt]
\Psi^{(0,4)} &=& \psi^{(0,4)} -4 \dot{a} \left( B^{(0,3)}-\frac{a}{2}\dot{E}^{(0,2)}\right)  + \frac{1}{2}\left(\nabla^{-2}\chi_{L ij}^{(0,4),ij} -\chi^{(0,4)}_{L}\right)  \\[5pt]
\mathbf{h}_{ij}^{(0,4)} &=& \hat{h}_{ij}^{(0,4)} + 2\chi^{(0,4)}_{Lij} +\left(\nabla^{-2}\chi_{Lkl}^{(0,4),kl} - \chi^{(0,4)}_L\right)\delta_{ij} + \nabla^{-2}\left(\nabla^{-2}\chi^{(0,4),kl}_{Lkl} + \chi^{(0,4)}_{L}\right)_{,ij} -4\nabla^{-2}\chi^{(0,4),k}_{Lk(i \;\;\; j)}  \, , \qquad 
\eea
where $\chi_{Lij}^{(0,4)}$ is defined such that
\bea
\chi_{Lij}^{(0,4)} &=& -\left(-\psi^{(0,2)}_{,k}\delta_{ij} + \frac{1}{2}E^{(0,2)}_{,ijk} + \frac{1}{2}F^{(0,2)}_{(i,j)k} + \frac{1}{2}\hat{h}^{(0,2)}_{ij,k} \right)\left(E^{(0,2),k} +F^{(0,2)k} \right) \nonumber \\[5pt]
&\quad & - \left(-\psi^{(0,2)}\delta_{ik} + \frac{1}{2}E^{(0,2)}_{,ik} + \frac{1}{2}F^{(0,2)}_{(i,k)} + \frac{1}{2}\hat{h}^{(0,2)}_{ik} \right)\left(E^{(0,2),k}_{,j} + F^{(0,2)k}_{,j}  \right) \nonumber \\[5pt]
&\quad & - \left(-\psi^{(0,2)}\delta_{jk} + \frac{1}{2}E^{(0,2)}_{,jk} + \frac{1}{2}F^{(0,2)}_{(j,k)} + \frac{1}{2}\hat{h}^{(0,2)}_{jk} \right)\left(E^{(0,2),k}_{,i} + F^{(0,2)k}_{,i}  \right) \, .  
\eea
This gives a full set of gauge invariant quantities for the post-Newtonian sector of our theory, up to the order that we are considering.

\vspace{10pt}
\noindent
{\bf Mixed-order quantities:}
Finally, at $\mathcal{O}(\epsilon \eta)$ we can construct two scalar and one tensor gauge invariant quantities:
\bea \label{eqMixedInvarPhi}
\Phi^{(1,1)} & =& \phi^{(1,1)} -\frac{1}{2} \phi^{(0,2)}_{,i} \left(E^{(1,0),i} + F^{(1,0)i}\right) \\
\Psi^{(1,1)} &=& \psi^{(1,1)}  + \frac{1}{2}\left(\nabla^{-2}\chi_{Lij}^{(1,1),ij} -\chi^{(1,1)}_L\right)   \\[5pt]
\mathbf{h}^{(1,1)}_{ij} &=&  \hat{h}^{(1,1)}_{ij} + 2 \chi^{(1,1)}_{Lij}  - 4\nabla^{-2}\chi^{(1,1),k}_{Lk(i,j)} + \nabla^{-2} \chi^{(1,1),kl}_{Lkl} \delta_{ij} - \chi^{(1,1)}_L \delta_{ij} 
   + \nabla^{-2}\nabla^{-2}\chi^{(1,1),kl}_{Lkl,ij} + \nabla^{-2}\chi^{(1,1)}_{L,ij} \, .
\eea
At order $\mathcal{O}(\epsilon \eta^2)$ there exists two scalar, one vector and one tensor gauge invariant quantities:
\bea \label{eqMixedInvarPhi}
\Phi^{(1,2)} & =& \phi^{(1,2)} + \dot{\phi}^{(0,2)}\left(aB^{(1,0)} -\frac{a^2}{2} \dot{E}^{(1,0)}\right)  + 2 \phi^{(0,2)}\left(\dot{a}B^{(1,0)} + a \dot{B}^{(1,0)} - a \dot{a}\dot{E}^{(1,0)} - \frac{a^2}{2} \ddot{E}^{(1,0)}\right)   \\[5pt]
\Psi^{(1,2)} &=& \psi^{(1,2)} + \nabla^{-2}\left(\chi^{(1,2),k]l}_{Lk[l} + 2 \mathcal{C}_{Lk[l\vert ,m}^{,\vert k]}\mathcal{I}^{m,l}_L\right)  \\[5pt]
\mathbf{B}^{(1,2)}_i &=& B^{(1,2)}_ i - \frac{a}{2} \dot{F}^{(1,1)}_i +\chi^{(1,2)}_{Li} -\nabla^{-2}\chi^{(1,2),j}_{Lj,i}      \\[5pt]
\mathbf{h}^{(1,2)}_{ij} &=& \hat{h}^{(1,2)}_{ij} + 2 \chi^{(1,2)}_{Lij}  - 4\nabla^{-2}\chi^{(1,2),k}_{Lk(i,j)} + \nabla^{-2} \chi^{(1,2),kl}_{Lkl} \delta_{ij} - \chi^{(1,2)}_L \delta_{ij} 
   + \nabla^{-2}\nabla^{-2}\chi^{(1,2),kl}_{Lkl,ij} + \nabla^{-2}\chi^{(1,2)}_{L,ij} \nonumber \\ 
   && + 4 \nabla^{-2} \nabla^{-2} \left(  \nabla^2 \mathcal{C}_{Lij,mk} \mathcal{I}^{m,k}_L - \nabla^2 \C_{Lk(i,j)m}\I^{m,k}_L -2\C_{Lk(i,j)klm}\I^{m,l}_L -\nabla^2 \C_{Lk(i \vert, m}^{,k} \I^m_{L,\vert j)} + \C^{,k( l \vert }_{Lkl,mn}\I^{m, \vert n)}_L\delta_{ij}   \right) \nonumber \\
   && +\nabla^{-2} \nabla^{-2}\left( - \nabla^2 \C^k_{Lk,ml} \I^{m,l}_L \delta_{ij} +2\C_{Lkl,mij}^{,k}\I^{m,l}_L + 2 \C^{,kl}_{Lkl,m(i} \I^m_{L,j)} + 2\C_{Lij,mk} \I^{m,k}_L      \right)\, .
\eea
The definitions of $\chi^{(1,2)}_{Li}$, $\chi^{(1,2)}_{Lij}$ and $\chi^{(1,1)}_{Lij}$ are given by
\bea
\chi^{(1,2)}_{Li} &=& \phi^{(0,2)}\left(B^{(1,0)} -\frac{a}{2}\dot{E}^{(1,0)}\right)_{,i} -\frac{a}{2} \left(-\psi^{(0,2)}\delta_{ij} +\frac{1}{2}E^{(0,2)}_{,ij} +\frac{1}{2}F^{(0,2)}_{(i,j)} + \frac{1}{2}\hat{h}^{(0,2)}_{ij} \right)\left(E^{(1,0),j} +F^{(1,0)j} \right)\dot{•}   \nonumber \\
&& -\frac{1}{2} \left(\frac{1}{2}B^{(0,3)}_{,i}+B^{(0,3)}_{i} -\frac{a}{4} \dot{F}^{(0,2)}_i   \right)_{,j} \left(E^{(1,0),j} + F^{(1,0)j}\right) \nonumber \\
&& -\frac{1}{2} \left(\frac{1}{2}B^{(1,0)}_{,j}+B^{(1,0)}_{j} -\frac{a}{4}\dot{F}^{(1,0)}_i \right) \left(E^{(0,2),j} + \delta F^{(0,2)j}\right)_{,i} \\
\chi^{(1,2)}_{Lij} &=& a\left(-\psi^{(0,2)} \delta_{ij} + \frac{1}{2}E^{(0,2)}_{,ij} + \frac{1}{2}F^{(0,2)}_{(i,j)} + \frac{1}{2}\hat{h}^{(0,2)}_{ij}  \right)\dot{•} \, \left(B^{(1,0)} -\frac{a}{2}\dot{E}^{(1,0)} \right) \nonumber \\[5pt]
&\quad & + 2\dot{a} \left( -\psi^{(0,2)} \delta_{ij} + \frac{1}{2}\hat{h}^{(0,2)}_{ij} \right)\left(B^{(1,0)} -\frac{a}{2}\dot{E}^{(1,0)} \right) \nonumber \\[5pt]
&\quad & -\frac{1}{2} \left(-\psi^{(0,2)} \delta_{ik} + \frac{1}{2}E^{(0,2)}_{,ik} + \frac{1}{2}F^{(0,2)}_{(i,k)} + \frac{1}{2}\hat{h}^{(0,2)}_{ik} \right)\left(E^{(1,0),k} +F^{(1,0)k} \right)_{,j} \nonumber \\[5pt]
&\quad & -\frac{1}{2} \left(-\psi^{(0,2)} \delta_{jk} + \frac{1}{2}E^{(0,2)}_{,jk} + \frac{1}{2}F^{(0,2)}_{(j,k)} + \frac{1}{2}\hat{h}^{(0,2)}_{jk} \right)\left(E^{(1,0),k} + F^{(1,0)k} \right)_{,i} \nonumber \\[5pt]
&\quad & -\frac{1}{2} \left(-\psi^{(1,0)} \delta_{ik} + \frac{1}{2}E^{(1,0)}_{,ik} + \frac{1}{2}F^{(1,0)}_{(i,k)} + \frac{1}{2}\hat{h}^{(1,0)}_{ik} \right)\left(E^{(0,2),k} + F^{(0,2)k} \right)_{,j} \nonumber \\[5pt]
&\quad & -\frac{1}{2} \left(-\psi^{(1,0)} \delta_{jk} + \frac{1}{2}E^{(1,0)}_{,jk} + \frac{1}{2}F^{(1,0)}_{(j,k)} + \frac{1}{2}\hat{h}^{(1,0)}_{jk} \right)\left(E^{(0,2),k} + F^{(0,2)k} \right)_{,i}    \\
\chi^{(1,1)}_{Lij} &=&  \C_{Lij,k} \I^{k}_L \, , 
\eea
where $\C_{Lij,k}$ and $\I^k_L$ are given by
\bea
\C_{Lij,k} & \equiv &   \left(-\psi^{(0,2)} \delta_{ij} + \frac{1}{2}E^{(0,2)}_{,ij} + \frac{1}{2}F^{(0,2)}_{(i,j)} + \frac{1}{2}\hat{h}^{(0,2)}_{ij} \right)_{,k}   \\
\I^k_L &\equiv &  -\frac{1}{2} \left(E^{(1,0),k} +F^{(1,0)k}\right) \, . \label{chiL11ij}
\eea 
This completes our study of gauge invariant quantities constructed from perturbations of the metric. 

It can be seen that there are a number of perturbed quantities in our formalism that are automatically gauge-invariant. These include the scalar Newtonian and post-Newtonian potentials $\phi^{(0,2)}$ and $\psi^{(0,2)}$, as well as the lowest-order tensor perturbations $\hat{h}_{ij}^{(1,0)}$ and $\hat{h}_{ij}^{(0,2)}$. The first two are expected as (depending on how one writes the field equations) they correspond to the gravitational potential in the Newton-Poisson equation. The last two show that the leading-order transverse and trace-free perturbations are invariant in both sectors of the theory. Comparing the form of the gauge-invariant quantities $\Phi^{(1,0)}$ and $\Phi^{(0,4)}$, it is interesting to note that they differ by a single term: $-\frac{1}{2}\phi^{(0,2)},_i(E^{(0,2),i}+F^{(0,2)i})$, which is quadratic in perturbations. The cosmological gauge invariant quantity $\Phi^{(1,0)}$ cannot contain a term of this form, as it would be higher order, at $\mathcal{O}(\epsilon^2)$. A number of other terms can be seen to occur in more than one of our gauge invariant quantities, and demonstrates the effect that the different length scales have on the order of perturbed quantities.

\subsection{Gauge invariant quantities from the energy-momentum tensor}

Let us now consider how to construct gauge invariant quantities from perturbations of the energy-momentum tensor. Again, our gauge invariant quantities will reduce to sources of stress-energy in the longitudinal gauge when $E=B=F_i=0$. We will do this first for the cosmological sector, and then for the post-Newtonian sector.

\vspace{10pt}
\noindent
{\bf Cosmological and mixed-order quantities:} We can construct the following three gauge-invariant scalars, corresponding to the mixed-order and cosmological energy density and pressure:
\bea
\mathbf{\rho}^{(1,1)} &=& \rho^{(1,1)} -\frac{1}{2}\rho^{(0,2)}_{,i}\left(E^{(1,0),i} + F^{(1,0)i}\right) \\
\mathbf{\rho}^{(1,0)} + \mathbf{\rho}^{(1,2)} &=&  \rho^{(1,0)} + \rho^{(1,2)}  + \dot{\rho}^{(0,2)} \left(aB^{(1,0)} - \frac{a^2}{2}\dot{E}^{(1,0)}\right)  \\[5pt]
\mathbf{p}^{(1,0)} &=& p^{(1,0)} \, .
\eea
The reader may note that $\rho^{(1,0)} + \rho^{(1,2)}$ transform together and give quadratic terms. They transform together because $\rho^{(1,0)}$ and $\rho^{(1,2)}$ are of the same order, $\mathcal{O}(\epsilon \eta^2 L_N^{-2})$, in our framework, even though $\rho^{(1,0)}$ is the leading-order large-scale perturbation to the energy density.

One further scalar, $\mathbf{v}^{(1,0)}$, and a divergence-free vector, $\mathbf{\hat{v}}^{(1,0)}_i$, can be extracted from the following gauge invariant quantity: 
\bea
\mathbf{v}_i^{(1,0)} &\equiv& \mathbf{v}^{(1,0)},_i + \mathbf{\hat{v}}^{(1,0)}_i = v_i^{(1,0)} + \frac{a}{2}\left(\dot{E}^{(1,0)}_{,i} + \dot{F}_i^{(1,0)} \right)   - \frac{1}{2}v^{(0,1)}_{i,j}\left(E^{(1,0),j} + F^{(1,0)j}\right) \, , \label{V10invar} 
\eea
by simply taking the divergence of it. These are all of the gauge invariant quantities that can be constructed from the energy-momentum tensor, in the cosmological and mixed-order sector of our theory.

\vspace{10pt}
\noindent
{\bf Post-Newtonian quantities:} In the post-Newtonian sector we have, at $\mathcal{O}(\eta)$, the following gauge-invariant quantities:
\bea
\mathbf{v}^{(0,1)} &=& v^{(0,1)}   \\[5pt]
\mathbf{\hat{v}}^{(0,1)}_i &=& \hat{v}^{(0,1)}_{i} \, ,
\eea
which we use to define the gauge invariant velocity $\mathbf{v}^{(0,1)}_i \equiv \mathbf{v}^{(0,1)}_{,i} +\mathbf{\hat{v}}^{(0,1)}_i$. At $\mathcal{O}(\eta^2)$ we find
\bea
\mathbf{\rho}^{(0,2)} &=&  \rho^{(0,2)} \, ,
\eea
and at $\mathcal{O}(\eta^4)$ we have
\bea
\mathbf{\rho}^{(0,4)} &=&  \rho^{(0,4)} - \rho^{(0,2)}\left(E^{(0,2),i} +F^{(0,2)i}\right) \\[5pt]
\mathbf{p}^{(0,4)} &=&  p^{(0,4)} \, .
\eea
This is again unsurprising, as many of these objects appear in the Newtonian equations of hydrodynamics. There are no further quantities in the energy-momentum tensor, so this gives us a full set of gauge invariant quantities in our 2-parameter perturbative expansion.

\subsection{Field equations in terms of gauge-invariant quantities} \label{sec:FinalFieldEqs}

With our newly-constructed gauge invariant quantities in hand, we can return to the field equations presented in Section \ref{sec:fe}. These equations take the same form as the field equations in the longitudinal gauge but are in fact valid in any coordinate system. Furthermore, these equations can be used to write down the governing equations for our gauge invariant quantities, which, upon specification of any particular gauge, should reduce to the gauge-fixed Einstein equations. As before, we write down these equations under the assumptions $\epsilon \sim \eta^2$ and $L_N/L_C \sim \eta$.

Note that we leave out both terms $R^{(2,0)}_{\mu \nu}$, in Eq. (\ref{anom1}), and $T^{(2,0)}_{\mu \nu}$ from the field equations. These terms appear in the $\mathcal{O}(\eta^4L_N^{-2})$ field equation as simply the lower order $00$-field equations $\mathcal{O}(\eta^2L_N^{-2})$ with two spatial derivatives and multiplied by two gauge generators, and so necessarily cancel and do not contribute any new dynamics to the field equations. 

\subsubsection{Background-order potentials}

The background-order $00$-field equation can be used to write
\bea
&\ & \frac{\ddot{a}}{a} + \frac{1}{6a^2}\nabla^2 \Phi^{(0,2)} = - \frac{4 \pi}{3}  {\mathbf \rho}^{(0,2)}\, , \label{FINAL0002} 
\eea
while the trace of the background-order $ij$-equation gives
\bea
&\ & \left( \frac{\dot{a}}{a} \right)^2 - \frac{1}{3 a^2} \nabla^2 \Phi^{(0,2)} = \frac{8 \pi}{3} {\mathbf \rho}^{(0,2)} \, ,
\label{FINALij02}
\eea
where we have substituted in the result that $\Phi^{(0,2)} = - \Psi^{(0,2)}$, found below in Eq. (\ref{psiphihij02}). The background order trace-free $ij$-equation gives
\bea
D_{ij}\left(\Phi^{(0,2)} + \Psi^{(0,2)} \right) - \frac{1}{2}\nabla^2 \mathbf{h}_{ij}^{(0,2)} =0 \, , 
\eea
and its derivative implies 
\be
\Phi^{(0,2)} = - \Psi^{(0,2)} \qquad {\rm and} \qquad \mathbf{h}^{(0,2)}_{ij}= 0 \, . \label{psiphihij02}
\ee
Note that all equations in this section are written with the substitution of the results in Eq. (\ref{psiphihij02}). The above equations govern the leading-order part of the gravitational field, at $\mathcal{O}(\eta^2L_N^{-2})$.

\subsubsection{Vector potentials}
We now use all $0i$-field equations. At order $\mathcal{O}(\eta^3L_N^{-2})$, these give
\bea
&\ & \nabla^2 {\mathbf B}^{(0,3)}_{i} + {2 }\left(a \dot{\Phi}^{(0,2)} + {\dot{a}}\Phi^{(0,2)}\right)_{,i} = 16 \pi a^2 {\mathbf \rho}^{(0,2)} {\mathbf v}^{(0,1)}_i \, .  \label{FINAL0i03} 
\eea
Although ${\mathbf B}^{(0,3)}_{i}$ is a divergenceless vector, Eq. (\ref{FINAL0i03}), has a divergenceless vector and scalar part, which can be separated out with a derivative, as can all equations in this section. At $\mathcal{O}(\eta^4L_N^{-2})$ the $0i$-field equations give
\bea
&&\nabla^2 \left({\mathbf B}^{(1,0)}_{i} +{\mathbf B}^{(1,2)}_{i}\right) +2 \left(a \left(\Phi^{(1,1)}-\Psi^{(1,0)}\right)\dot{•}+ \dot{a}\left(\Phi^{(1,1)}+ \Phi^{(1,0)} \right)   \right)_{,i}
 -2 \left( 2 {\dot{a}^2} + a {\ddot{a}} \right) {\mathbf B}^{(1,0)}_{i} - {\mathbf B}^{(1,0)}_{j}\Phi^{(0,2)}_{,ij}\nonumber \\
&&= 8 \pi a^2 \left( 2 \mathbf{\rho}^{(1,1)}\mathbf{v}^{(0,1)}_i +{\mathbf \rho}^{(0,2)}\left( {\mathbf B}^{(1,0)}_i + 2 {\mathbf v}^{(1,0)}_i \right) \right) \, .
\label{FINAL0i04}
\eea
We note that the vector part of Eq. (\ref{FINAL0i04}) is not sourced by quadratic lower-order potentials, although at first glance it looks like it may be.

\subsubsection{Higher-order scalar potentials}
The $00$- and $ij$-trace field equation at $\mathcal{O}(\epsilon \eta L_N^{-2})$ give 
\bea
&\ &   \frac{1}{6a^2}\nabla^2 \Phi^{(1,1)} = - \frac{4 \pi}{3}  {\mathbf \rho}^{(1,1)}\, , \label{FINAL0011} 
\eea
and imply
\be
\Phi^{(1,1)} = - \Psi^{(1,1)} \, . \label{condition11} 
\ee

Using the $00$-field equation, at $\mathcal{O}(\eta^4L_N^{-2})$, we find 
\bea
&& \nabla^2 \left( \Phi^{(1,0)} + \frac{1}{2}\Phi^{(0,4)} + \Phi^{(1,2)}\right) + \left(\nabla \Phi^{(0,2)}\right)^2
+{3 a \dot{a}}\left(3\Phi^{(0,2)} + \Phi^{(1,0)} -2 \Psi^{(1,0)}\right)\dot{•} \nonumber \\[5pt]
&& + {3 a^2}\left(\Phi^{(0,2)}- \Psi^{(1,0)}\right)\, \ddot{•} 
 + 6 a {\ddot{a}}\left(\Phi^{(0,2)} - \Psi^{(1,0)}\right)  - \frac{1}{2} \Phi^{(0,2)}_{,ij}\mathbf{h}^{(1,0)}_{ij} \nonumber \\[5pt]
&=&-8 \pi a^2 \left( {\mathbf \rho}^{(1,0)} + {\mathbf \rho}^{(1,2)} + \frac{1}{2} {\mathbf \rho}^{(0,4)} + 3\left(\mathbf{p}^{(1,0)} + \mathbf{p}^{(0,4)}\right)   - {\mathbf \rho}^{(0,2)} \left( \Phi^{(1,0)} + \Psi^{(1,0)} - 2 \left(\mathbf{v}^{(0,1)}_i\right)^2  \right)   \right) . 
\label{FINAL0004} 
\eea
The trace of the $ij$-field equation gives, at $\mathcal{O}(\eta^4L_N^{-2})$,
\bea
&& - 2 \nabla^2 \left(\Psi^{(1,0)} + \Psi^{(1,2)} + \frac{1}{2}\Psi^{(0,4)}\right) 
- 3\left(2 \dot{a}^2 + a \ddot{a}\right)\left(\Phi^{(1,0)} - \Psi^{(1,0)} + 2\Phi^{(0,2)}\right) 
+ 6\dot{a}a\left(\Psi^{(1,0)} -\Phi^{(0,2)} \right)\dot{}
 \nonumber \\[5pt]
&=& - 4 \pi a^2 \left( 4\left({\mathbf \rho}^{(1,0)} + {\mathbf \rho}^{(1,2)} + \frac{1}{2} {\mathbf \rho}^{(0,4)}\right) 
+ {\mathbf \rho}^{(0,2)} \left( 2\Phi^{(0,2)} - \Phi^{(1,0)} -3\Psi^{(1,0)} + 4 \left(\mathbf{v}^{(0,1)}_i\right)^2 \right) \right) + \mathcal{A} \, , \label{FINALijtrace04}
\eea
where we have defined terms that are quadratic in metric potentials as 
\bea
\mathcal{A} &\equiv& \nabla^2 \Phi^{(0,2)}\left(3\Phi^{(0,2)} + \frac{1}{2}\Phi^{(1,0)} - \frac{5}{2}\Psi^{(1,0)}\right) + \frac{3}{2}\left(\nabla \Phi^{(0,2)}\right)^2 + \frac{1}{2}\Phi^{(0,2)}_{,ij}\mathbf{h}^{(1,0)}_{ij} \, . 
\label{FINALA}
\eea
These are all of the scalar equations that exist at this order.

\subsubsection{Tensor potentials}

The trace-free $ij$-field $\mathcal{O}(\epsilon \eta L_N^{-2})$ equation is
\bea
D_{ij}\left(\Phi^{(1,1)} + \Psi^{(1,1)} \right) - \frac{1}{2}\nabla^2 \mathbf{h}_{ij}^{(1,1)} =0 \, , \label{FINALij11}
\eea
and its derivative implies 
\be
\Phi^{(1,1)} = - \Psi^{(1,1)} \qquad {\rm and} \qquad \mathbf{h}^{(1,1)}_{ij}= 0 \, . \label{psiphihij11}
\ee
However, note that unlike $\Psi^{(0,2)}$ and $\Phi^{(0,2)}$, the condition that $\Phi^{(1,1)} = - \Psi^{(1,1)}$ is already given by the $00-$ and $ij-$trace field equations, Eq. (\ref{FINAL0011}), that imply Eq. (\ref{condition11}). We substitute the results in Eq. (\ref{psiphihij11}) into all equations in this section. Finally, the $ij$-field equation, at $\mathcal{O}(\eta^4L_N^{-2})$, can be used to write the following trace-free equation:
\bea
&& - D_{ij}\left(\Phi^{(1,0)} + \Phi^{(1,2)} + \frac{1}{2}\Phi^{(0,4)} + \Psi^{(1,0)} + \Psi^{(1,2)} + \frac{1}{2}\Psi^{(0,4)}\right) 
 + \frac{1}{2} \nabla^2 \left(\mathbf{h}_{ij}^{(1,0)} + \mathbf{h}_{ij}^{(1,2)} + \frac{1}{2}\mathbf{h}_{ij}^{(0,4)}\right) \nonumber \\[5pt]
&& +4 \dot{a}\left(\mathbf{B}_{(i,j)}^{(0,3)} + \mathbf{B}_{(i,j)}^{(1,0)}\right) 
+2 a\left(\mathbf{B}_{(i,j)}^{(0,3)} + \mathbf{B}_{(i,j)}^{(1,0)}\right)\dot{•} - \left(2\dot{a}^2 + a\ddot{a}\right)\mathbf{h}^{(1,0)}_{ij} 
 -\frac{3}{2} a\dot{a}\dot{\mathbf{h}}_{ij}^{(1,0)} - \frac{1}{2} a^2  \ddot{\mathbf{h}}_{ij}^{(1,0)}  \nonumber \\[5pt]
& =& -8 \pi a^2 {\mathbf \rho}^{(0,2)} \left(\frac{1}{2}\mathbf{h}^{(1,0)}_{ij} + 2 \mathbf{v}^{(0,1)}_{\langle i} \mathbf{v}^{(0,1)}_{j \rangle}\right) + \mathcal{B}_{ij} \, ,
\label{FINALijtracefree04}
\eea
where we have defined terms that are quadratic in metric potentials as 
\bea
\mathcal{B}_{ij} &\equiv& D_{ij}\Phi^{(0,2)}\left(2\Phi^{(0,2)} + \Phi^{(1,0)} -\Psi^{(1,0)}\right)  
+  \Phi^{(0,2)}_{,\langle i}\Phi^{(0,2)}_{, j \rangle} -  \Phi^{(0,2)}_{,k \langle i }\mathbf{h}^{(1,0)}_{j\rangle k} \, .
\label{FINALBij}
\eea
We observe that, unlike in linear cosmological perturbation theory, our expansion scheme does not imply $\Phi^{(1,0)} = -\Psi^{(1,0)}$ or $\mathbf{h}_{ij}^{(1,0)} =0$ because of the additional potentials in Eq. (\ref{FINALijtracefree04}) that do not exist in cosmological perturbation theory. This completes the full set of equations for our gauge-invariant variables, up to the order in perturbations that we wish to consider here.

\end{widetext}

\section{Discussion} \label{SmallAndLarge}

Using our two-parameter expansion we will now discuss the application of it to various physical situations that are of interest. Note that although Sections \ref{sec:LargeLim} and \ref{SSlimit} consider post-Newtonian structure on very different scales, as do all systems considered in Section \ref{obsjust}, gravitational potentials remain small and of similar size $\epsilon \sim \eta^2$.

\subsection{Large-scale limit: $l \sim \eta$} \label{sec:LargeLim}

Let us now discuss the field equations given in Section \ref{sec:FinalFieldEqs}. In this case the small-scale structure is on the scale of superclusters, $L_N \sim 100$Mpc so $l \sim \eta$, and gravitational potentials are such that $\epsilon \sim \eta^2$ (as justified in Section \ref{obsjust}). Firstly, we note that in the lowest-order field equations, (\ref{FINAL0002}) and (\ref{FINALij02}), the Newtonian mass density and gravitational potentials source the evolution of the scale factor. In the next-to-leading-order field equations, (\ref{FINAL0i03}), (\ref{FINAL0011}) and (\ref{FINALij11}), we have mixed-order and post-Newtonian potentials, but no quadratic source terms, meaning that these field equations are not sourced by the lowest-order field equations. In the $\mathcal{O}(\eta^4)$ field equations, Eqs. (\ref{FINAL0i04}), (\ref{FINAL0004}), (\ref{FINALijtrace04}) and (\ref{FINALijtracefree04}), on the other hand, we find first order cosmological, mixed-order, Newtonian and post-Newtonian potentials. This means that linear-order cosmological perturbations (that usually arise as first-order corrections to the background field equations) in fact come in after two lower order field equations. In addition, the $\mathcal{O}(\eta^4)$ field equations that contain the linear-order cosmological potentials are sourced by quadratic lower-order potentials. These effects only arise because of the form of our two-parameter expansion, and so do not (and cannot) occur in linear-order cosmological perturbation theory. 

The reader may note that our expansion requires field equations to exist at orders that simply do not exist in cosmological perturbation theory. For example, in cosmological perturbation theory the leading-order vector mode (which contributes to frame-dragging effects) decays quickly, and so is usually taken to be zero. However, the magnitude of the second-order part of this potential has recently been found to be much bigger than one might naively estimate -- between $\mathcal{O}(\epsilon)$ and $\mathcal{O}(\epsilon^2)$ \cite{AndEtAl}, at about $O(\epsilon^{1.5})$. In our expansion we already have a vector potential at order $\eta^3 \sim \epsilon^{1.5}$, and it is clear that such a potential should exist from the post-Newtonian perturbed sector. This means that the result of Ref. \cite{AndEtAl}, which look a little odd in the context of cosmological perturbation theory, fit very naturally into our framework. Our expansion also suggests that there should be field equations at $\mathcal{O}(\eta^5)$, which would correspond to a potential of $\mathcal{O}(\epsilon^{1.5})$ in normal cosmological perturbation theory. This simply does not exist in the usual expansion, but is included if one follows the approach we have used in this paper.

The reader may also note that cosmological perturbation theory is not recovered by simply setting $\eta \to 0$. This is because in cosmological perturbation theory the lowest order energy density is homogeneous, whereas in the late Universe, as described by our two-parameter expansion, the lowest order energy density is inhomogeneous (see Section \ref{pert2para}). We therefore cannot recover cosmological perturbation theory by ignoring the post-Newtonian sources, as when $\eta \to 0$ the evolution of the scale factor in Eq. (\ref{FINAL0002}) would have no source at all. This means that post-Newtonian sector must be included, in both the equations for the background expansion and the linear-order cosmological perturbations. Specifically, this means that standard cosmological perturbation theory is not necessarily recovered if one averages over some length scale greater than or equal to the homogeneity scale, as is usually assumed \cite{chrisBack}. To compare our two-parameter expansion to cosmological perturbation theory we must average the field equation (\ref{FINALij02}) over a suitably large scale. 

We start by calculating the average energy density $\overline{\rho}$, obtained from integrating over volumes, $V_{\mathrm{hom}}$, that correspond to the homogeneity length scale, $L_{\mathrm{hom}} \sim 100\mathrm{Mpc}$ \cite{Hogg}. This gives 
\bea
\overline{ \mathbf{\rho}}^{(0,2)} &\equiv & \frac{\int_{V_{\mathrm{hom}}} \mathbf{\rho}^{(0,2)} dV }{\int_{V_{\mathrm{hom}}} dV} \, . \label{rhobar2}
\eea
The closest thing we can then define to the usual first-order part of the energy density, $\delta \mathbf{\rho}$, is then
\bea
\delta \mathbf{\rho}^{(0,2)}  & \equiv & \mathbf{\rho}^{(0,2)}  -  \overline{\mathbf{\rho}}^{(0,2)} \, .
\eea
This means that the leading-order inhomogeneous part of the energy density, $\delta \mathbf{\rho}$, is the same order as the background, $\overline{\mathbf{\rho}}$, both being $O (\eta^2 L_N^{-2})$. Finally, one may note that derivatives of $\delta \mathbf{\rho}$ go like $1/L_N$, and not $1/L_C$.

Let us now outline how to start solving the field equations (\ref{FINAL0002})-(\ref{FINALBij}). We first take the lowest order field equation, given by Eq. (\ref{FINALij02}), and integrate this over the volume corresponding to the homogeneity scale
\begin{equation} \label{intefe}
-\frac{1}{a^2}\int_{V_{\textrm{hom}}} \nabla^2 \Phi^{(0,2)} dV + \int_{V_{\textrm{hom}}} 3 H^2 dV = \kappa \int_{V_{\textrm{hom}}} \mathbf{\rho}^{(0,2)} \, , 
\end{equation}
where $\dot{a}/a \equiv H$ and $\kappa = 8 \pi$. Using Gauss' theorem this can be written
\bea \label{intefe2}
-\frac{1}{a^2}\int_{S_{\textrm{hom}}} \nabla \Phi^{(0,2)} \cdot dS + 3H^2 V_{\textrm{hom}} = \kappa M^{(0,2)},
\eea
where $M^{(0,2)}$ is the total rest mass in the volume $V_{\rm hom}$. If we now assume that on the homogeneity scale there is no net flux of $\nabla \Phi^{(0,2)}$ into or out of the surface $S_{\textrm{hom}}$, then the first term in Eq. (\ref{intefe2}) vanishes. This leaves us with
\begin{equation}
3H^2 = \kappa \overline{\mathbf{\rho}}^{(0,2)} \, , \label{homolowest}
\end{equation}
where, from Eq. (\ref{rhobar2}), $\overline{\mathbf{\rho}}^{(0,2)} = {M^{(0,2)}}/{V_{\textrm{hom}}}$. Finally, substituting these results into Eq. (\ref{FINALij02}) gives
\begin{equation} 
\nabla^2 \Phi^{(0,2)} = - \kappa a^2 \delta \mathbf{\rho}^{(0,2)} \, . \label{inhomolowest}
\end{equation}
This equation can be solved using Green's functions, N-body simulations or Fourier methods. Moreover, it provides justification for why it is only the average energy density that sources the large-scale expansion, while it is the energy density minus its average that sources the Newton-Poisson equation, even though both Eqs. (\ref{homolowest}) and (\ref{inhomolowest}) are of the same order. The key here is the existence of a homogeneity scale at which there is no net flux in $\nabla \Phi^{(0,2)}$, which seems like a restrictive but necessary condition in order to derive Eqs. (\ref{homolowest}) and (\ref{inhomolowest}). It means that for the system to be perturbed FLRW globally we need matter to be strictly distributed such that the average energy density in {\it every} region is the same.

Finally, we comment that our two-parameter expansion was constructed such that perturbations on scales above the cut-off of 100Mpc are treated as cosmological, whereas perturbations below this cut-off are treated as post-Newtonian. This cut-off is somewhat artificial. In the real Universe there are structures, such as Baryon Acoustic Osccilations, that exist on approximately the scale of this cut-off. The practical application of our two-parameter expansion to model such structures would require further thought, and perhaps some flexibility. 

\subsection{Small-scale limit: $l \ll \eta$} \label{SSlimit}

Let us consider what would happen if we considered structure on the smallest scales, similar to the solar system for example, such that $L_N \sim L_{\odot} \ll  \eta L_C$. The first thing to happen would be that long-wavelength cosmological perturbations in the energy density, $\rho^{(1,0)}$ for example, would be relegated to very high-order field equations compared to those presented in Section \ref{sec:FinalFieldEqs}, because $L_{\odot} \ll \eta L_C \ll L_C$. Moreover, the `post-Newtonian' order energy density would be replaced by ${\textstyle\frac{1}{2}} \rho^{(0,4)}  + \rho^{(1,2)}$. To disentangle $\rho^{(1,2)}$ and $\rho^{(0,4)}$ one would then have to use the fact that $\rho^{(1,2)}$ has large-scale correlations, whereas $\rho^{(0,4)}$ does not. The reader may also note that if $l \ll \eta$ then this implies there is no $\rho^{(1,1)}$, $h^{(1,1)}_{00}$ or $h^{(1,1)}_{ij}$ for that matter (see Section \ref{pert2para}). 

However, there does remain a potential $h^{(1,2)}_{0i}$, which appears in the field equations at $\mathcal{O}(\eta^4)$ if $\epsilon \sim \eta^2$. This does not occur in usual post-Newtonian gravity, where the $0i$-field equations contain terms at $\mathcal{O}(\eta^3)$ and then at $\mathcal{O}(\eta^5)$. This means that the mixed term $h^{(1,2)}_{0i}$ would correspond to a $\eta^4$ correction to the post-Newtonian $\eta^3$ $0i$-field equation. Nevertheless, $h^{(1,2)}_{0i} \sim \eta^4$ is at higher order than anything that has so far been observed in the solar system\footnote{The best observational constraints on this potential have made up to an accuracy of about 20\% accuracy with Gravity Probe B's gyroscope precession experiment \cite{GPB}, and about 5\% accuracy with the LAGEOS and LARES satellites \cite{LAGEOS}.}, as current observations have only allowed the $0i$ metric potential to be constrained to $\mathcal{O}(\eta^3)$. Our formalism is therefore consistent with observed post-Newtonian gravity to date, but may offer a new opportunity to test gravity at higher orders in the future, as more accurate observations may one day be able to detect gravitational phenomena associated with $h^{(1,2)}_{0i}$. 

Finally, if $l \ll \eta$ then the field equations will be dominated by the Newton-Poisson equation at lowest order. Cosmological terms such as $\ddot{a} \sim 1/L_C^2$ and $ \nabla^2 h^{(1,0)}_{00} \sim \epsilon/L_C^2$ will only occur at much higher order. Although the leading-order parts of post-Newtonian gravity and our two-parameter expansion are indistinguishable when applied to structure on small scales, at higher orders (or for structures on larger scales) our formalism also includes terms that account for the sourcing of the expansion of the scale factor and large-scale cosmological potentials. These corrections simply do not appear in the usual approach to post-Newtonian gravity, where cosmological expansion is entirely neglected. However, the reader may also note that we recover the usual post-Newtonian expansion in the limits $\epsilon \to 0$ and $a(t) \to 1$. 

\subsection{Other systems}

Let us now consider other scenarios that one might try to model with a two-parameter approach of the type described in this paper, that do not fall into the two cases described above, or may not satisfy $\epsilon \sim \eta^2$. The first thing that one may note for such a situation is that our two-parameter expansion simply does not allow for post-Newtonian-perturbed structures larger than the supercluster scale of $100$Mpc, so great walls or voids larger than this scale cannot be considered within this expansion (see Eq. (\ref{xleqeta})). If such situations were considered, then the lowest order field equation would be $\ddot{a} = 0$, which only has the solutions $a \propto t$. We note that for post-Newtonian perturbed structures smaller than supercluster scales $l < \eta$ the field equations will behave similarly to those discussed in Section \ref{SSlimit}, specifically the scale factor would be sourced at higher order, as would all terms with derivatives or units $L_C$, and Newtonian gravity would dominate.

Now consider cases where $\epsilon > \eta^2$. This could be the case, for example, in a universe full of low-mass stars or high density contrast voids. In this case and for $l \sim \eta$ the evolution of the scale factor would remain in the lowest order field equation, at $\mathcal{O}(\eta^2 L_N^{-2})$, with the energy density. Long-wavelength cosmological perturbations, on the other hand, would be squeezed in somewhere between the lowest Newtonian order, $\mathcal{O}(\eta^2 L_N^{-2})$, and first post-Newtonian order, $\mathcal{O}(\eta^4 L_N^{-2})$, for $00$- and $ij$- field equations. Nevertheless, by construction, the cosmological energy density must be strictly less than Newtonian one (see Eq. (\ref{left})). 

Finally, if $\eta^2 > \epsilon$ then the expansion around FLRW is still valid but may start to break down if $\eta \rightarrow 1$. This would be the case close to compact objects, such as neutron stars and black holes. In these cases cosmological perturbations are relegated to higher order. Of course, in the real Universe these strong gravity scenarios tend to happen on small-scales, when $L_N \ll \eta L_C$. In these cases we would expect the scale factor to be sourced at higher order too.

As a last remark, if one were to consider a system with structure on more than two scales, say $N$ scales, an $N$-parameter expansion would probably be necessary. Nevertheless, structure on supercluster scales would always remain the dominant contributor to the expansion of the scale factor, as discussed throughout this section. 

\section{Conclusion} \label{conc}

We propose and constructed a two-parameter perturbative expansion around an FLRW metric that can simultaneously describe non-linear structures on small scales, and linear structures on large scales. We find that the gravitational potentials from small-scale structures can source the growth of structure on large scales, and that one should in general expect mode mixing in the equations that govern the large-scale fluctuations. The effects are of significance observationally, as the next generation of surveys will be able to measure fluctuations in the density contrast on scales approaching the entire observable Universe. Understanding the behaviour of these fluctuations in the presence of non-linear structure is of importance not only for removing potential sources of bias, but also because it has the potential to offer new ways of looking for the effects of Einstein's theory. This could come about through the generation of non-Gaussianity, through the form of the matter power spectrum on large scales, or the identification of novel new effects that do not occur in linearised gravity. We consider our perturbative expansion to contain some of the essential features of the real late Universe, and therefore to have a number of potential advantages over standard cosmological perturbation theory.

The work we have presented in this paper contains a derivation of the field equations, an explicit presentation of a two-parameter gauge transformation, and the construction of gauge invariant quantities in both the matter and gravity sectors of the theory. We find that consistency of the gauge transformations requires not only gravitational potentials and matter perturbations at the orders expected from post-Newtonian gravity and cosmological perturbation theory, but also a number of others at orders of perturbation where they may not naively have been expected. We have therefore identified a minimal set of perturbations that are required for mathematical consistency of the problem, and written down gauge invariant versions of the field equations that contain them all. 

We discuss the application of our formalism to a universe containing different gravitational systems. This includes a universe containing post-Newtonian structure on solar system scales, for which our field equations are consistent with post-Newtonian gravity up to the accuracy of current observations but differ at higher order. The field equations we derive account for structure on the scale of clusters and superclusters within the context of cosmological perturbations, and we find that, with a certain notion of homogeneity above scales of around $100$Mpc, it is possible to write down a version of the Friedmann equation in which the expansion is driven by the average rest mass density, from the post-Newtonian sector of the theory. The small-scale Newton-Poisson equations for the scalar gravitational potentials occur at the same order in perturbations as the Friedmann equation, while the lowest order equations that contain the cosmological gravitational potentials appear at higher order. These latter equations contain post-Newtonian matter sources, and quadratic Newtonian-level potentials from small scales. They therefore contain valuable information about non-linear gravity, and could potentially be used to identify relativistic effects in large-scale structure observations.

\section*{Acknowledgements}

We are grateful to P. Carrilho and J. C. Hidalgo for helpful discussions and comments. SRG, KAM and TC acknowledge support from the STFC. The tensor algebra package xAct \cite{xAct} and its sub-package xPand \cite{xPand1, xPand2}, were used to derive some of the equations presented in this work.

\section*{APPENDIX: PERTURBED RICCI AND ENERGY-MOMENTUM TENSORS} \label{appendixlong}

This appendix provides detailed expressions for the perturbed Ricci tensor and the perturbed energy-momentum tensor, which are used to derive the field equations presented in Section \ref{sec:fe}. We make no assumptions about the relative magnitude of $\epsilon$ and $\eta$ in this appendix, nor do we assume anything about the length scales $L_C$ and $L_N$.

We begin by expanding the components of the Ricci tensor in our two parameters. We find that the non-vanishing contributions to each component are given by the following equations:
\bea
R_{00} &=& R_{00}^{(0,0)} + R^{(0,2)}_{00} + R^{(0,3)}_{00}\label{R00pert}\\[5pt]&&
+ {\textstyle\frac{1}{2}}R^{(0,4)}_{00}+ R_{00}^{(1,0)}  + R^{(1,1)}_{00}+ R^{(1,2)}_{00} + \ldots   \nonumber 
\\[5pt]&& \nonumber \\[5pt]
R_{0i} &=&  R_{0i}^{(0,2)}+ R_{0i}^{(0,3)}+R_{0i}^{(1,0)}+ R_{0i}^{(1,2)} + \ldots \label{R0ipert} 
\\[5pt]&& \nonumber \\[5pt]
R_{ij} &=& R_{ij}^{(0,0)} + R^{(0,2)}_{ij} + R^{(0,3)}_{ij}
\label{Rijpert} \\[5pt]&& + {\textstyle\frac{1}{2}}R^{(0,4)}_{ij}+ R_{ij}^{(1,0)} + R^{(1,1)}_{ij}+ R^{(1,2)}_{ij} + \ldots \, , \nonumber
\eea
where ellipses denote higher-order terms, which we will not require in this paper. 

Any term in each of these equations has an order of smallness in $\epsilon$ and $\eta$, as indicated by the superscript in brackets. They also have a length scale associated with them, given by $L_N^{-2}$, $L_C^{-2}$ or $L_C^{-1} L_N^{-1}$. We have not indicated this directly on each of the terms in the expansion, but it is important when using these equations to determine the field equations presented in Section \ref{sec:fe}. We will therefore be careful to keep track of them in the expressions that follow.

The terms on the right-hand side of Eq. (\ref{R00pert}) are given explicitly by
\bea
R_{00}^{(0,0)} &=& - 3 \frac{\ddot{a}}{a} \sim \frac{1}{L_C^2} \label{R0000} \\[5pt] 
R_{00}^{(0,2)} &=& - \frac{1}{2a^2}h^{(0,2)}_{00,ii}  \sim \frac{\eta^2}{L_N^2} \label{R0002} \\[5pt]
R_{00}^{(0,3)} &=& \frac{\dot{a}}{a^2}h^{(0,3)}_{0i,i} -\frac{\dot{a}}{a}h^{(0,2)}_{ii,0} 
- \frac{3\dot{a}}{2a}h^{(0,2)}_{00,0} \sim \frac{\eta^3}{L_C L_N} \;\;\;\;\;\;  \label{R0003}\\[5pt]
R_{00}^{(0,4)} &=& -\frac{1}{2 a^2} \left( h_{00,i}^{(0,2)} \right)^2 - \frac{1}{2a^2}h^{(0,4)}_{00,ii} - h^{(0,2)}_{ii,00} \label{R0004} \\[5pt] &\quad & + \frac{2}{a}h^{(0,3)}_{0i,0i} + \frac{1}{2a^2} h^{(0,2)}_{00,i} \left( 2h^{(0,2)}_{ij,j} - h^{(0,2)}_{jj,i} \right) \nonumber \\[5pt] &\quad & + \frac{1}{a^2}h^{(0,2)}_{00,ij}h^{(0,2)}_{ij} \sim  \frac{\eta^4}{L_N^2} \nonumber\\[5pt]
R_{00}^{(1,0)} &=& - \frac{1}{2a^2} h^{(1,0)}_{00,ii} - \frac{1}{2}h^{(1,0)}_{ii,00} + \frac{\dot{a}}{a^2} h^{(1,0)}_{0i,i} \label{R0010} \\[5pt] &\quad &- \frac{\dot{a}}{a} h^{(1,0)}_{ii,0} + \frac{1}{a}h_{0i,0i}^{(1,0)} - \frac{3\dot{a}}{2 a}h^{(1,0)}_{00,0}\sim  \frac{\epsilon}{L_C^2} \nonumber \\[5pt]
R_{00}^{(1,1)} &=& -\frac{1}{2a^2}h^{(1,1)}_{00,ii} \sim \frac{\epsilon \eta}{L_N^2} \label{R0011} \\[5pt]
R_{00}^{(1,2)} &=& -\frac{1}{2a^2}h^{(1,2)}_{00,ii} +\frac{1}{2a^2}h^{(0,2)}_{00,ij}h^{(1,0)}_{ij} \label{R0012} \\[5pt]
&\quad &+ \,  \mathrm{terms \, of \, size} \, \left[ \frac{\epsilon \eta^2}{L_N L_C} \right] \nonumber \\[5pt] 
&\sim & \frac{\epsilon \eta^2}{L_N^2} +\frac{\epsilon \eta^2}{L_N L_C}\, \nonumber . 
\eea
The terms in Eq. (\ref{R0ipert}) are given by
\bea 
R_{0i}^{(0,2)} &=&- \frac{\dot{a}}{a} h^{(0,2)}_{00,i} \sim \frac{\eta^2}{L_C L_N} 
\label{R0i02}\\[5pt]
R_{0i}^{(0,3)} &=& \frac{1}{2a} \left( h^{(0,3)}_{0j,ij} -h^{(0,3)}_{0i,jj} +a h^{(0,2)}_{ij,0j} -a h^{(0,2)}_{jj,0i} \right)  \label{R0i03}\\[5pt]
&\quad & + \,  \mathrm{terms \, of \, size} \, \left[ \frac{\epsilon \eta^3}{L_C^2} \right] \nonumber \\[5pt]  
&\sim &   \frac{\eta^3}{L_N^2} +\frac{\eta^3}{L_C^2} \nonumber\\[5pt]
R_{0i}^{(1,0)} &=& \frac{1}{2a}\Big( h^{(1,0)}_{0j,ij} - h^{(1,0)}_{0i,jj} +ah^{(1,0)}_{ij,0j} -ah^{(1,0)}_{jj,0i} \label{R0i10} 
\\[5pt] && -2\dot{a}h^{(1,0)}_{00,i} + 4\dot{a}^2h^{(1,0)}_{0i} + 2a\ddot{a} h^{(1,0)}_{0i}  \Big) \nonumber \sim  \frac{\epsilon}{L_C^2} \\[5pt]
R_{0i}^{(1,1)}&=& -2\dot{a}h^{(1,1)}_{00,i} \sim \frac{\epsilon\eta}{L_NL_C} \\[5pt]
R_{0i}^{(1,2)} &=& \frac{1}{2a}\left( h^{(1,2)}_{0j,ij} - h^{(1,2)}_{0i,jj} +ah^{(1,1)}_{ij,0j} -ah^{(1,1)}_{jj,0i} \label{R0i12}
  \right)\\[5pt] &&+ \frac{1}{2a}h^{(1,0)}_{0j}h^{(0,2)}_{00,ij} \nonumber \\[5pt]
&\quad & + \,  \mathrm{terms \, of \, size} \, \left[ \frac{\epsilon \eta^2}{L_C^2} + \frac{\epsilon \eta^2}{L_NL_C} \right] \nonumber \\[5pt] 
&\sim & \frac{\epsilon \eta^2}{L_N^2} + \frac{\epsilon \eta^2}{L_C^2} + \frac{\epsilon \eta^2}{L_NL_C}\, .\nonumber 
\eea
Finally, the terms in Eq. (\ref{Rijpert}) are given by
\bea 
R_{ij}^{(0,0)} &=& \left( 2\dot{a}^2 + a\ddot{a} \right) \delta_{ij} \sim \frac{1}{L_C^2}, \label{Rij00} \\[5pt]
R_{ij}^{(0,2)} &=&  \frac{1}{2} \left( h^{(0,2)}_{00,ij} + 2h^{(0,2)}_{k(i,j)k} - h^{(0,2)}_{kk,ij} - h^{(0,2)}_{ij,kk} \right)  \label{Rij02}\\[5pt] 
&&+\left (2\dot{a}^2+ a \ddot{a} \right) \left( h^{(0,2)}_{ij} +h^{(0,2)}_{00} \delta_{ij} \right)  \nonumber\\[5pt]
&&\sim \frac{\eta^2}{L_N^2} +\frac{\eta^2}{L_C^2} \nonumber \\[5pt]
R_{ij}^{(0,3)} &=&\frac{1}{2}a \dot{a}h^{(0,2)}_{00,0}\delta_{ij} - 2 \dot{a}h^{(0,3)}_{0(i,j)} - \dot{a}h^{(0,3)}_{0k,k}\delta_{ij} \label{Rij03}\\[5pt]
&&+  \frac{3}{2}a \dot{a}h^{(0,2)}_{ij,0} + \frac{1}{2}a \dot{a}h^{(0,2)}_{kk,0}\delta_{ij}  \sim  \frac{\eta^3}{L_C L_N} \nonumber\\[5pt]
R_{ij}^{(0,4)} &=& \frac{1}{2} \left( h^{(0,4)}_{00,ij} -h^{(0,4)}_{ij,kk} -h^{(0,4)}_{kk,ij} \right)  + h^{(0,4)}_{k(i,j)k} \label{Rij04} \\[5pt] 
&& + a^2 h^{(0,2)}_{ij,00} + \frac{1}{2} h^{(0,2)}_{00,k} \left( h^{(0,2)}_{ij,k} -2h^{(0,2)}_{k(i,j)} \right) \nonumber \\[5pt] &\quad & + h^{(0,2)}_{kl,ij}h^{(0,2)}_{kl} +  h^{(0,2)}_{ij,kl}h^{(0,2)}_{kl}  -2h^{(0,2)}_{k(i,j)l}h^{(0,2)}_{kl} \nonumber \\[5pt] &\quad &+ \frac{1}{2} h^{(0,2)}_{kl,i}h^{(0,2)}_{kl,j}  + h^{(0,2)}_{kl,l} \left( h^{(0,2)}_{ij,k} - 2h^{(0,2)}_{k(i,j)} \right)  \nonumber \\[5pt] &\quad &+  h^{(0,2)}_{ik,l} \left( h^{(0,2)}_{jk,l} -h^{(0,2)}_{jl,k} \right) + \frac{1}{2} h^{(0,2)}_{00,i}h^{(0,2)}_{00,j} \nonumber \\[5pt] &&     \nonumber +  h^{(0,2)}_{00,ij}h^{(0,2)}_{00} +   h^{(0,2)}_{kk,l} \left( 2h^{(0,2)}_{l(i,j)}  -  h^{(0,2)}_{ij,l} \right) \nonumber \\[5pt]
&& -2a h^{(0,3)}_{0(i,j)0}  + \,  \mathrm{terms \, of \, size} \, \left[ \frac{\eta^4}{L_C^2} \right] \nonumber \\[5pt] &\sim &  \frac{\eta^4}{L_N^2}  + \frac{\eta^4}{L_C^2}\nonumber \\[5pt]
R_{ij}^{(1,0)} &=& \frac{1}{2} \left( h^{(1,0)}_{00,ij} -h^{(1,0)}_{ij,kk} -h^{(1,0)}_{kk,ij} \right) + h^{(1,0)}_{k(i,j)k} \label{Rij10} \\[5pt] 
&&+ a \ddot{a}h^{(1,0)}_{ij}  + a \ddot{a} h^{(1,0)}_{00} \delta_{ij} + 2 \dot{a}^2h^{(1,0)}_{00} \delta_{ij} \nonumber \\[5pt] && + \frac{1}{2} a \dot{a} h^{(1,0)}_{00,0}\delta_{ij} -2 \dot{a}h^{(1,0)}_{0(i,j)} -\dot{a}h^{(1,0)}_{0k,k} \delta_{ij}   \nonumber \\[5pt] && + \frac{3}{2} a \dot{a} h^{(1,0)}_{ij,0}  + \frac{1}{2} a \dot{a}h^{(1,0)}_{kk,0}\delta_{ij} + \frac{1}{2}a^2 h^{(1,0)}_{ij,00} 
\nonumber \\[5pt]
&&+ 2\dot{a}^2h^{(1,0)}_{ij}- a h^{(1,0)}_{0(i,j)0} \sim \frac{\epsilon}{L_C^2} \nonumber \\[5pt]
R_{ij}^{(1,1)} &=&  \frac{1}{2}(h^{(1,1)}_{00,ij} -h^{(1,1)}_{ij,kk} -h^{(1,1)}_{kk,ij} ) +h^{(1,1)}_{k(i,j)k} \label{Rij11} \\[5pt]
&\quad & + \,  \mathrm{terms \, of \, size} \, \left[ \frac{\epsilon \eta}{L_C^2} \right] \nonumber \\[5pt]
&\sim & \frac{\epsilon \eta}{L_N^2} + \frac{\epsilon \eta}{L_C^2}  \nonumber \\
R_{ij}^{(1,2)} &=&  \frac{1}{2}\left(h^{(1,2)}_{00,ij} -h^{(1,2)}_{ij,kk} -h^{(1,2)}_{kk,ij} \right) +h^{(1,2)}_{k(i,j)k} \label{Rij12} \\[5pt] &\quad & + \frac{1}{2}h^{(0,2)}_{00,ij}h^{(1,0)}_{00}   + \frac{1}{2}h^{(0,2)}_{kl,ij}h^{(1,0)}_{kl}+ \frac{1}{2}h^{(0,2)}_{ij,kl}h^{(1,0)}_{kl} \nonumber \\[5pt] &\quad &   - h^{(0,2)}_{k(i,j)l}h^{(1,0)}_{kl} \nonumber \\[5pt]
&\quad & + \,  \mathrm{terms \, of \, size} \, \left[ \frac{\epsilon \eta^2}{L_C^2} + \frac{\epsilon \eta^2}{L_NL_C} \right] \nonumber \\[5pt]
&\sim & \frac{\epsilon \eta^2}{L_N^2} + \frac{\epsilon \eta^2}{L_C^2} + \frac{\epsilon \eta^2}{L_NL_C}\, , \nonumber
\eea
where in Eq. (\ref{Rij02}) the two orders or magnitude after the $\sim$ indicate the first and second lines, respectively.

Let us now consider the energy-momentum tensor, $T_{\mu \nu}$. Expanding in both $\epsilon$ and $\eta$ the non-vanishing components of this tensor are given by 
\bea 
T_{00} &=& T^{(0,2)}_{00}  + T^{(0,4)}_{00}+ T_{00}^{(1,0)} + T^{(1,1)}_{00}+ T^{(1,2)}_{00} + \ldots \qquad \quad \label{T00pert} \\[5pt] \nonumber \\[5pt]
T_{0i} &=&  T_{0i}^{(0,3)} + T_{0i}^{(1,2)} + \ldots \label{T0ipert} \\[5pt] \nonumber \\[5pt]
T_{ij} &=&  T^{(0,4)}_{ij} + T^{(1,0)} \ldots \, , \label{Tijpert}
\eea
where ellipses again indicate higher-order terms that we will not consider in this study. The terms on the right-hand side of Eq. (\ref{T00pert}) are given by
\bea
T^{(0,2)}_{00} &=&  \rho^{(0,2)} \sim \frac{\eta^2}{L_N^2}  \label{T0002} \\[5pt]
T^{(0,4)}_{00} &=& \frac{1}{2} \rho^{(0,4)} - h^{(0,2)}_{00}\rho^{(0,2)} \label{T0004} \\[5pt] && +  \rho^{(0,2)} v^{(0,1)i}v^{(0,1)}_i\sim  \frac{\eta^4}{L_N^2} \nonumber  \\[5pt]
T^{(1,0)}_{00} &=&  \rho^{(1,0)} \sim \frac{\epsilon}{L_C^2} \label{T0010} \\[5pt]
T^{(1,1)}_{00} &=&  \rho^{(1,1)} \sim \frac{\epsilon \eta}{L_N^2}  \label{T0011} \\[5pt]
T^{(1,2)}_{00} &=&  \rho^{(1,2)} - h^{(1,0)}_{00}\rho^{(0,2)}     \sim \frac{\epsilon \eta^2}{L_N^2} \, , \label{T0012} 
\eea
while the terms in Eq. (\ref{T0ipert}) are given by
\bea
T^{(0,3)}_{0i} &=&  -a \rho^{(0,2)}v_i^{(0,1)} \sim \frac{\eta^3}{L_N^2} \label{T0i03} \\[5pt]
T^{(1,2)}_{0i} &=&  -a\left(\rho^{(0,2)}v^{(1,0)}_i +\rho^{(1,1)}v^{(0,1)}_i\right) \label{T0i12}  \\[5pt]
&& - a \rho^{(0,2)}h^{(1,0)}_{0i}  + \, \mathrm{terms \, of \, size \, } \left[ \frac{\epsilon \eta^2}{L_C^2} \right]  \nonumber \\[5pt]
& \sim& \frac{\epsilon \eta^2}{L_N^2} + \frac{\epsilon \eta^2}{L_C^2}\, , \nonumber \;\;\;\; 
\eea
and the terms in Eq. (\ref{Tijpert}) are given by
\bea
T^{(0,4)}_{ij} &=&   a^2 \rho^{(0,2)}v^{(0,1)}_i v^{(0,1)}_j + a^2 p^{(0,4)} \delta_{ij} \sim \frac{\eta^4}{L_N^2} \; \; \; \; \label{Tij04} \\[5pt]
T^{(1,0)}_{ij} &=&   a^2 p^{(1,0)} \delta_{ij} \sim \frac{\epsilon}{L_C^2}\, . \label{Tij10}
\eea
This completes the list of expanded tensor components that are required for Section \ref{sec:fe}.

\end{document}